\documentclass{optica-article}

\journal{opticajournal} % for journals or Optica Open

\articletype{Research Article}

\usepackage{array} % for big tables with fixed widths
\usepackage{rotating}
\usepackage{pdflscape}
\usepackage{hyperref}

\begin{document}

\title{Practical Limits on Integrated Squeezers}

\author{
Devin~J.~Dean,\authormark{1} 
Taewon~Park,\authormark{1,2}
Lars~S.~Madsen,\authormark{3,4}
Alex~Terrasson,\authormark{3,4}
Sam~Robison,\authormark{1}
Geun~Ho~Ahn,\authormark{1,2}
Ziyu~Wang,\authormark{1}
Hubert~S.~Stokowski,\authormark{1}
Luke~Qi,\authormark{1}
Jesse~J.~Slim,\authormark{3,4}
Joel~Corney,\authormark{4}
Darwin~Serkland,\authormark{5}
Warwick~P.~Bowen,\authormark{3,4}
Martin~M.~Fejer,\authormark{1}
Amir~H.~Safavi-Naeini\authormark{1,*}
}

\address{%
\authormark{1}Department of Applied Physics and Ginzton Laboratory, Stanford University, Stanford, California, USA\\

\authormark{2}Department of Electrical Engineering, Stanford University, Stanford, California, USA\\

\authormark{3}ARC Centre of Excellence in Quantum Biotechnology, St Lucia, Queensland, Australia\\

\authormark{4}School of Mathematics and Physics, University of Queensland, St Lucia, Queensland, Australia\\

\authormark{5}Sandia National Laboratories, Albuquerque, New Mexico, USA
}%

\email{\authormark{*}safavi@stanford.edu} %% email address is required; see note below about the corresponding author designation

% use {asbstract*} to suppress the copyright line. Copyright information will be added in production

\begin{abstract*} 
Recent experiments have demonstrated the successful generation and detection of moderately squeezed vacuum states with integrated photonics. However, in order to benefit from the reduced noise of highly squeezed light, many different noise sources must be mitigated. Here, we quantify the fundamental limits these noise sources impose on squeezing measurements and find surprising generality across different platforms and designs. We combine these different limitations into a simple model that provides practical guidance for the design and benchmarking of next-generation integrated squeezed-light systems.

\end{abstract*}

%%%%%%%%%%%%%%%%%%%%%%%%%%%%%%%%%%%%%%%%%%%%%%%%%%%%%%%%%%%%%%%%%%%%%%%%%

% \tableofcontents

\section*{Introduction}\label{Sec:Intro}
%%%%%%%%%%%%%%%%%%%%%%%%%%%%%%%%%%%%%%%%%%%%%%%%%%%%%%%%%%%%%%%%%%%%%%%%%

%%%%%%%%%%%%%%%%%%%%%%%%%%%%%%%%%%%

Laser light has quantum fluctuations. Squeezed light is a kind of light whose quantum fluctuations are redistributed such that one quadrature has less noise than typical while the other has more. The unique noise properties and quantum correlations of squeezed light makes it potentially useful in various fields ranging from optical sensing \cite{LIGO2020, Bowen2021} to quantum computation \cite{Madsen2022}.
Squeezed light was first successfully measured in 1985 by Richart Slusher and colleagues \cite{PhysRevLett.55.2409} using resonant four-wave mixing, and shortly thereafter by Ling-An Wu and colleagues~\cite{KimbleWu1986} using resonant three-wave mixing. It has been used to enhance the sensitivity of the Laser Interferometer Gravitational-Wave Observatory (LIGO) since 2019 \cite{LIGO2019}, and is now routinely generated in experimental laboratories around the world. In the journey to measuring highly squeezed light, many unexpected noise sources were encountered and had to be overcome \cite{Levenson1986}, resulting in the now-substantial literature on the original squeezer design -- a bulk signal-resonant squeezer\cite{Lam1999, Chelkowski_2007, Mavalvala2014}. 

The development of integrated nonlinear photonic circuits provides us with the opportunity to develop squeezed light technologies, transitioning from pioneering experiments that primarily used bulk optics to compact, scalable, and low-power chip-scale systems. Recent experiments in integrated nonlinear photonic circuits have measured modest amounts of squeezing, mainly limited by excess losses \cite{hirota2025, Furusawa2023, Kashiwazaki2021, Inoue2023, Nehra2022, Shi2025, Park2024, Andersen2024, Stokowski2023, Ulanov2025, Vernon2021, Gaeta2020, Politi2020, Purdy2013, Safavi-Naeini2013}.
To our knowledge, there is no unified model that brings together the relevant noise sources for these next-generation squeezers.

This paper is intended to serve as a guide to the design space of integrated squeezed light sources, providing a summary of the potential limiting factors for realistic devices and combining them in a simple analytic model. We hope that this formalized model will make it easier to compare different squeezer designs and material platforms and quickly estimate limiting factors. 

\begin{figure}
    \centering
    \includegraphics[width=\linewidth]{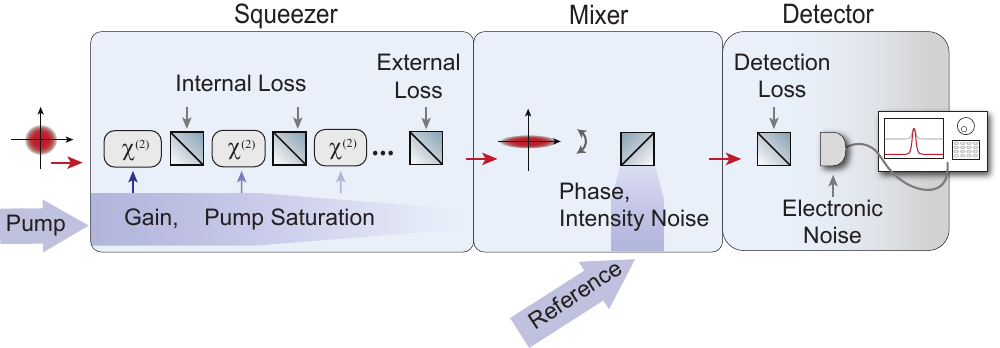}
    \caption{\label{fig:sqz_concept} \textbf{Noise sources in squeezing experiments.} A generic squeezing measurement consists of a squeezer, a mixer, and a detector. Each section has different noise contributions that can limit the total measured squeezing. Red circle on the left represents vacuum noise in phase space, which is transformed by the squeezer into the squeezed ellipse shown in the mixer section, which after detection is shown by the reduced noise on the electronic spectrum analyzer. $\chi^{(2)}$ nonlinear sections are depicted within the squeezer for concreteness - some squeezers instead utilize $\chi^{(3)}$ nonlinear sections. The squares with diagonal slashes represent the effective beamsplitters that model how many processes mix in noise. 
 }
\end{figure}

A main result of our work is a compact equation that collects the most important physical noise processes into a single variance budget. Although many previous studies have focused on individual effects in isolation, here we group them into two natural categories: terms associated with the \textit{generation} of squeezed light and terms associated with its \textit{detection}. In our model, depicted conceptually in Fig.~\ref{fig:sqz_concept}, the squeezed quadrature variance is approximately:
\begin{align}
    \label{eq:noises_sum}
V^- &\approx \underbrace{\big(V_\text{gain} + \Delta V_\text{internal loss}  + \Delta V_\text{pump saturation} + \Delta V_\text{external loss} + \Delta V_\text{phase noise} 
\big)}_{\text{generation}} \nonumber \\&+ \underbrace{\big(   \Delta V_\text{intensity noise} + \Delta V_\text{detection loss}   +  \Delta V_\text{electronic noise}  \big)}_{\text{detection}}.
\end{align}
The first group represents the limits set by nonlinear gain, internal propagation loss, pump saturation,  out-coupling, and imperfect phase control, and the second group represents the limits imposed by excess reference beam fluctuations, finite detector efficiency, and electronic noise.
The above expression is the backbone for the rest of the paper: in the following sections, we estimate the contributions individually and show how each depends on device and system parameters in the context of integrated devices.

%%%%%%%%%%%%%%%%%%%%%%%%%%%%%%%%%%%%%%%%%%%%%%%%%%%%
\section*{Results}
\subsection*{Theoretical Framework - Linearized Quantum States}

Our approach utilizes the framework of Gaussian quantum optics, where the noise associated with a quantum state is described by the variances in the $\hat X$ and $\hat Y$ quadratures - $V^X$ and $V^Y$.
See the Methods %\section \ref{sec:methods_gaussian} 
for more details on this framework of linearized quantum states.

For coherent states, $V^X = V^Y = 1$, which sets the so-called shot noise limit (SNL). A squeezed state is a state whose variance in one quadrature is below the SNL ($V^- < 1$). The quadrature variances are subject to the Heisenberg uncertainty product 
\begin{equation} \label{eq:Heisenberg}
V^X V^Y \ge 1,
\end{equation}
which implies that the other quadrature of a squeezed state must be antisqueezed ($V^+ > 1$).
 Going forward $V^\pm$ refer to the variance of the antisqueezed and squeezed quadratures, respectively. The reduced variance of squeezed states can be useful in optical sensors as it increases their signal-to-noise ratio.

In both squeezed light generation and detection, noise and loss processes can be modeled as beamsplitter interactions that mix the signal with a noise bath. The variance $V$ evolves according to
\begin{equation}
V_{\text{out}} = \eta V_{\text{in}} + (1-\eta)V_{\rm noise},
\end{equation}
where $\eta$ is the transmission efficiency ($0 \leq \eta \leq 1$) and $V_{\rm noise}$ is the variance of the added noise (typically vacuum fluctuations), which is assumed to be uncorrelated with the signal. For a sequence of $j$ such processes, the final variance $V_j$ is obtained by applying this relation sequentially. 
We show in the Methods %section \ref{sec:App:lowlossapprox} 
that, in the typical regime of small noise mixing $1-\eta_j \ll 1$, this equation simplifies to a simple sum
\begin{align}
    V_\text{out} 
    &\approx V_0 + (1-\eta_1)V_{\text{noise},1} + (1-\eta_2)V_{\text{noise},2} +\cdots +  (1-\eta_j)V_{\text{noise},j} \\
    &\approx V_0 + \Delta V_\text{noise,1} + \Delta V_\text{noise,2} + \cdots +  \Delta V_{\text{noise},j} .
\end{align} We show in the Methods that this approximation is valid for any system that measures at least 2 dB of squeezing, and becomes more exact the greater the squeezing. This simple sum is the basis for the variance budget presented in Eq. \eqref{eq:noises_sum}. To easily compare different noise contributions, we report the noise contributions from all processes in shot noise units.

%%%%%%%%%%%%%%%%%%%%%%%%%%%%%%%%%%%%%%%%%%%%%%%%%%%%%%%%%%%%%%%%%%%%%%%%%
\subsection*{Squeezed Light Generation}
%%%%%%%%%%%%%%%%%%%%%%%%%%%%%%%%%%%%%%%%%%%%%%%%%%%%%%

%%%%%%%%%%%%%%%%%%%%%%%%%%%%%%%%%%%%%%%%%%%%%%%%%%%%%%%

Squeezed light can be generated in degenerate $\chi^{(2)}$ optical parametric amplifiers, where a strong pump beam travels through a nonlinear crystal and generates squeezing at the signal frequency, which is half the frequency of the pump. 
Integrated devices have generated squeezing using nonlinear processes other than $\chi^{(2)}$ - such as by four wave mixing in $\chi^{(3)}$-nonlinear resonators~\cite{Gaeta2020, Vernon2021, Politi2020, Ulanov2025} and ponderomotive squeezing in optomechanical systems~\cite{Safavi-Naeini2013,Purdy2013}. Most of the following analysis is general to \textit{any} nonlinear optical squeezer in the Gaussian state approximation. Nonetheless, for concreteness, we consider a continuous-wave $\chi^{(2)}$ degenerate squeezer when necessary - mainly in Section "Pump Saturation - Nonresonant" %\ref{sec:Saturation_nonres} 
where we derive pump saturation effects in a $\chi^{(2)}$ squeezer.

Degenerate squeezing is  generated by a Hamiltonian that generates correlated pairs of photons in the same mode, and is given by 
\begin{equation}
    \frac{\hat H}{\hbar}=\frac{ig_t}{2} \hat a^2 - \frac{ig_t^\ast}{2} \hat a^{\dagger 2} ,
\end{equation}
where $g_t$ is the temporal nonlinear gain, and has units of $[\mathrm{time}^{-1}]$. Often, in nonlinear optics literature, it is more convenient to describe interactions in terms of spatially varying time-harmonic fields and we therefore describe the nonlinearity in terms of spatial gain rate $g$ with units $[\mathrm{length}^{-1}]$. These two are related by the group velocity $v_g$:  $g = g_t/ v_g$. We will refer mostly to the spatial nonlinear gain rate in future equations, in order to simplify the relation between device length and gain. In a $\chi^{(2)}$ squeezer, the spatial nonlinear gain rate is given by $g \equiv \sqrt{\eta_0 P},$ where $\eta_0$ is the normalized nonlinear efficiency (typically given in units $\%/\text{W~cm}^2$) and $P$ is the pump power \cite{Lvovsky2015}.

%%%%%%%%%%%%%%%%%%%%

The performance of a generic optical squeezer is given by a combination of gain, external loss, internal loss, pump depletion, and phase noise. We now discuss each of these factors individually and find general expressions that apply across different squeezer designs - unifying the limitations of the original signal-resonant squeezers with other designs, such as nonresonant and pump-resonant squeezers.

%%%%%%%%%%%%%%%%%%%%%%%%%%%%%%%%%%%%%%%%%%%%%%%%%%%%%%%
\subsubsection*{Gain} \label{sec:gain}
%%%%%%%%%%%%%%%%%%%%%%%%%%%%%%%%%%%%%%%%%%%%%%%%%%%%%%%

In a squeezer, one quadrature experiences deamplification (squeezing) while the other experiences amplification (gain). The amount of phase sensitive amplification defines the squeezer gain
\begin{equation}
G \equiv \frac{V^+}{V^+(0)}.
\end{equation}
where $V^+(0), V^+$ are the initial and final antisqueezed variances. Since the ideal squeezing interaction is unitary, it preserves the phase space area and therefore the squeezed quadrature variance is reduced by at most the same factor that the anti-squeezed quadrature is amplified. The gain limitation to noise variance, for use in Eq. \eqref{eq:noises_sum}, is therefore given by 
\begin{equation} \label{eq:V_gain}
\boxed{ V_\text{gain} = \frac{V^-(0)}{G}
},
\end{equation}
where $V^-(0)$ is the initial variance in the squeezed quadrature (unity for vacuum or coherent states).
High gain is therefore critical for generating high levels of squeezing. 

The squeezer gain depends on the spatial nonlinear gain rate $g$ and the interaction length $L$, both of which are constrained in practice.
Gain rate depends on pump power, which is limited by available laser power and the material's optical damage threshold, and the normalized efficiency, which is determined by the material nonlinearity and mode confinement. 
Although interaction length is not limited by diffraction in waveguide squeezers as it is in bulk squeezers, it is still limited by pump propagation loss, phase mismatch, and footprint constraints. 
In practice, resonant cavities are often used to effectively enhance both the gain rate and length. 

\begin{figure}
    \centering
    \includegraphics[width=1\linewidth]{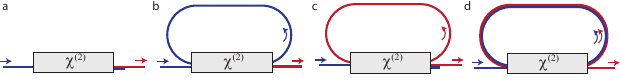}
    \caption{\label{fig:cavity_sqz} \textbf{Resonant squeezers.} Pump light (blue) is input to a nonlinear crystal and generates squeezing at the signal wavelength (red). (a) nonresonant (b) pump resonant (c) signal resonant (OPO) (d) fully resonant. }
\end{figure}

The gain of a nonresonant squeezer, shown in Figure \ref{fig:cavity_sqz}a where a traveling wave pump squeezes a traveling wave signal, is simply
\begin{equation} \label{eq:G_nonres}
    G_\text{nonres} = e^{2 g L}.
\end{equation}
In a pump-resonant squeezer, the pump is recirculated as in Figure \ref{fig:cavity_sqz}b and builds to higher powers that help it squeeze traveling wave light at the signal wavelength - the gain is still given by equation \eqref{eq:G_nonres} but with enhanced gain rate since $g \propto \sqrt P$ for $\chi^{(2)}$ squeezers. It is important to minimize reflections in high-gain nonresonant and pump-resonant squeezers, as even a small amount of feedback of the signal can lead to instability in the presence of large gain.

In a low-loss signal-resonant squeezer, depicted in Figure \ref{fig:cavity_sqz}c, light at the signal wavelength passes through the gain region many times and effectively extends the interaction length and enhances the gain
\begin{align} 
    G_\text{signal-res}= 1+\rho \frac{4 {g}/{g_{\mathrm{th}}}}{(1-g/g_{\mathrm{th}})^2}, \label{eq_OPO_Vp}
\end{align}
where the escape efficiency $\rho = 1-\frac{\alpha}{g_{th}}$ is the ratio of outcoupling rate to total cavity loss rate (or equivalently, one minus the ratio of loss rate to threshold gain rate),
$g$ is the same gain rate as before, and the threshold gain rate $g_{\mathrm{th}}$ is the gain rate above which optical parametric oscillation occurs - the round-trip gain exceeds round-trip signal loss and the initial vacuum fluctuations are amplified to macroscopic levels~\cite{CollettGardiner_1984, CollettGardiner_1985, Chelkowski_2007}. 
Signal-resonant squeezers must be operated below this oscillation threshold in order to produce squeezed vacuum, and only produce substantial squeezing within the cavity linewidth.
As threshold is approached $g \rightarrow g_{\mathrm{th}}$, the effective length and gain approach infinity. 
 The benefits of signal-resonant and pump-resonant squeezers can be combined in fully-resonant squeezers, as shown in Figure \ref{fig:cavity_sqz}d, albeit with increased complexity since all beams must be stable on resonance. 

The original continuous-wave bulk squeezers relied on signal resonance to enhance nonlinearity and enforce single-mode operation \cite{ Lam1999, Chelkowski_2007, Mavalvala2014, Schnabel2016}.
Conversely, emerging integrated platforms with strong single-pass nonlinearity have demonstrated large gains with only pump-resonance~\cite{Dean2026} or no resonance at all \cite{Furusawa2023}.
However, despite achieving large gains on all platforms, the achievable squeezing is still limited by other factors.

\subsubsection*{Loss}

Squeezed states are fragile in the presence of loss, which introduces vacuum noise. We define internal loss as loss within the squeezer, and external loss as loss that occurs after the squeezed state has been generated. External loss has limited most integrated squeezing measurements to date, and its vacuum noise contribution to squeezing is discussed in the following section. The contribution of internal loss is less well-known, and in this section we present a universal bound for internal loss that applies to all squeezer types.

\begin{figure}[h]
    \centering
    \includegraphics[width=\linewidth]{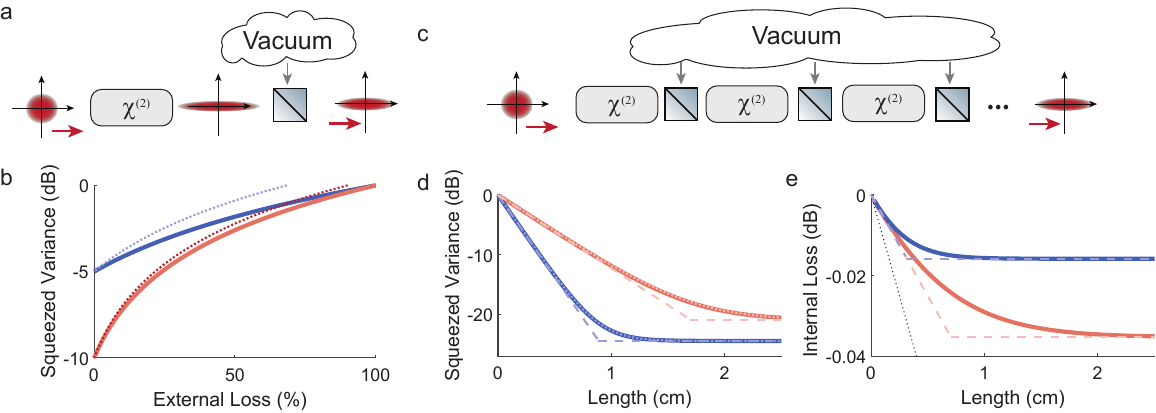}
    \caption{\label{fig:loss} \textbf{Effects of External and Internal Loss.} (a) External loss - loss after a squeezer introduces vacuum fluctuations, which reduces the squeezing as shown in the phase space diagrams. (b) Effect of external loss on squeezed variance (equation \eqref{eq:external_loss}), for input squeezing of 5 dB (blue) and 10 dB (red), with dotted lines representing simplified model of equation \eqref{eq:external_loss_approx}. (c) Internal Loss model of a lossy nonlinear region as a series of infinitesimal nonlinear regions and loss regions. 
    (d) Squeezing evolution with length for two gain rates ($g = 12$ dB/cm red and $g = 27$ dB/cm blue) and fixed loss rate $\alpha = 10$ dB/m for a nonresonant squeezer (equation \eqref{eq:internal_nonres}. Dashed lines represent the approximate limits of the two regimes of equation \eqref{eq:internal_regimes} for each gain rate, and dotted lines of the approximate model are indistinguishable from the exact solid lines.
    (e) Internal loss vs length for two gain rates and fixed loss rate $\alpha = 10$ dB/m, for a nonresonant squeezer. Dashed lines represent the approximate limits of the two regimes of equation \eqref{eq:nonres_internalloss_lims} for each gain rate. Dotted line represents total loss.
    }
\end{figure}

%%%%%%%%%%%%%%%%%%%%%%%%%%%%%%%%%%%%%%%%%%%%%%%%%%%%%%%
\paragraph{External Loss}
%%%%%%%%%%%%%%%%%%%%%%%%%%%%%%%%%%%%%%%%%%%%%%%%%%%%%%%

External loss after the squeezer degrades squeezing by the beamsplitter interaction
\begin{equation} \label{eq:external_loss}
V^- = \eta_e V^-(0) + (1-\eta_e),
\end{equation}
where $\eta_e$ is the fraction of light transmitted and the input squeezing is $V^-(0)$. Under the approximation of small external loss, the total squeezing becomes $V^- \approx V^-(0) + (1-\eta_e )$ and so we attribute the noise contribution due to the loss 
\begin{equation} \label{eq:external_loss_approx}
\boxed{ \Delta V_\text{external loss} = 1-\eta_e
}.
\end{equation}
The exact and approximate effects of external loss on squeezing are shown in Figure \ref{fig:loss}b, where the approximate equation \eqref{eq:external_loss_approx} is clearly valid for small losses and large initial squeezing. 
External loss presents a significant challenge in integrated squeezer systems, where the losses from separating the squeezed light from the pump beam, propagation, coupling between components, and the imperfect quantum efficiency of detectors all accumulate and make external loss one of the most common limiting factors for measured squeezing. 
Detection loss, or loss associated with coupling into the detector as well as the detector’s non-unity quantum efficiency, is discussed further in the "Squeezed Light Detection" part of this work.% Section \ref{Sec:DetectionLoss}.

%%%%%%%%%%%%%%%%%%%%%%%%%%%%%%%%%%%%%%%%%%%%%%%%%%%%%%%
\paragraph{Internal Loss}\label{Sec:internalloss}
%%%%%%%%%%%%%%%%%%%%%%%%%%%%%%%%%%%%%%%%%%%%%%%%%%%%%%%
Internal loss, arising from scattering and absorption within the squeezers, 
behaves differently than external loss due to the interplay of simultaneous squeezing and loss, as shown conceptually in Figure~\ref{fig:loss}c. Intuitively, loss near the end of the squeezer will degrade squeezing much more than loss near the start, since the output squeezed state is sensitive to loss but the input vacuum state is not. This is the opposite of how noise evolves in a typical amplifier, where the interplay of gain and loss make the noise figure more sensitive to loss at the input than at the output\cite{OPANFtheory}.
We quantify the internal loss by its effect on the squeezing separate from the gain $G$ discussed in the previous section:
we define an effective $\eta_i$ (distinct from the internal power transmission of the OPA) that describes the resultant reduction of the squeezing, given the gain $G$:
\begin{align} 
 V^- &= \eta_i \frac{1}{G} + (1-\eta_i),  \\ 
 \Delta V_\text{internal loss} &= 1-\eta_i =  \frac{ V^- - \frac{1}{G}}{1-\frac{1}{G} }  , \label{eq:internal_loss_def}
\end{align}
We will show in this section that the internal loss of all squeezer types is bounded by the inverse of nonlinearity-to-loss ratio $g/\alpha$
\begin{equation}
\boxed{
\Delta V_\text{internal loss} \gtrsim  \frac{\alpha}{g}
}.
\end{equation}
The ratio provides excellent intuition for the requirements of squeezers - in order to achieve 20 dB of squeezing given a loss rate $\alpha=0.1$ dB/cm, the gain rate must be at least a hundred times larger: $g> 10$ dB/cm. 
We will now derive simple expressions for the noise contributions from internal loss for the different squeezer designs. 

%%%%%%%%%%%%%%%%%%%%%%%%%%%%%
% \subsubsection*{Nonresonant and Pump-resonant}

The internal loss of a nonresonant squeezer was derived over 30 years ago (D. Serkland 1993, unpublished) by modeling the simultaneous loss and squeezing as a sequence of infinitesimal propagations through alternating regions of pure nonlinearity with spatial field gain rate $g$, and pure loss with intrinsic spatial field loss rate $\alpha$, as shown in Figure~\ref{fig:loss}c. We introduce this model in the Methods% \ref{app:internalloss}, 
assuming for simplicity that the spatial gain rate and loss rate are constant throughout the squeezer. 
The squeezed and antisqueezed quadrature variances evolve according to
\begin{align} \label{eq:internal_nonres}
V^- &= \left(V^-(0) - \frac{\alpha}{\alpha+g}\right) e^{-2(\alpha+g)L} + \frac{\alpha}{\alpha+g} \\
&\approx \frac{V^-(0)}{G} + \frac{\alpha}{\alpha + g},   &\  \frac{g}{\alpha} \gg 1 \label{eq:internal_regimes}\\ \nonumber
V^+ &=  \left(V^+(0) + \frac{\alpha}{g-\alpha}\right) e^{2(g-\alpha)L} -\frac{\alpha}{g- \alpha} \nonumber \\ 
& \approx GV^+(0), &\ \frac{g}{\alpha} \gg 1, \nonumber
\end{align}
where the approximation of large nonlinearity to loss ratio $\frac{g}{\alpha} \gg 1$ is valid for most squeezer systems.  
Figure~\ref{fig:loss}d plots equation \eqref{eq:internal_nonres} and shows that there are two distinct regimes of squeezing. Initially the squeezing evolves with gain as in the "Gain" Section %\ref{sec:gain} 
and is independent of the waveguide loss. However, after some propagation the squeezing reaches a minimum value set by the nonlinear and loss rates $V^-= \frac{\alpha}{\alpha + g}$. In this lossy squeezing regime, the anti-squeezed quadrature continues to grow and the state becomes a noisy non-minimum-uncertainty state \cite{Sipe2017, Yamashima:25_feedforward}.
The transition between lossless and lossy squeezing occurs at the transition length
\begin{equation}
L_0 = \frac{\ln(\frac{\alpha+g}{\alpha})}{2g}.
\end{equation}
The contribution of internal loss (equation \eqref{eq:internal_loss_def}) has two regimes:
\begin{align} \label{eq:nonres_internalloss_lims}
\Delta V_\text{internal loss, nonres}  &\approx \begin{cases}
 1-e^{-\alpha L} & L\leq L_{\eta_i}\\
 \frac{\alpha}{\alpha+g} & L>L_{\eta_i}  
\end{cases}
,  \ \ \ \ \text{ where } L_{\eta_i} \equiv \frac{1}{\alpha+g}
\end{align}
which are plotted in Figure \ref{fig:loss}e:
Initially it scales as the square root of the total loss. However, after some short propagation length the internal loss quickly saturates at the minimum value. The transition between these two internal loss regimes occurs at a length $L_{\eta_i}$  much shorter than the squeezing transition length $L_0$ we found earlier, so therefore for typical lengths we can approximate the internal loss as the (constant) second regime
\begin{equation}
\Delta V_\text{internal loss, nonres} \approx \frac{\alpha}{\alpha + g} .
\end{equation}
which in the usual limit of $g\gg\alpha$ reduces to the inverse of the nonlinearity to loss ratio $\frac{\alpha}{g}$. 
The internal loss of a pump-resonant squeezer is identical to that of a nonresonant squeezer, but with the gain rate resonantly enhanced due to the enhanced pump power.

%%%%%%%%%%%%%%%%%%%%%%%%%%%%%%%%%%%
% \subsubsection*{Signal Resonant}

The squeezing produced by a signal-resonant squeezer is \cite{Chelkowski_2007}
\begin{align} \label{eq:opo_V_m}
    V^- = 1-\rho \frac{4 {g}/{g_{\mathrm{th}}}}{(1+g/g_{\mathrm{th}})^2} ,
\end{align}
which accounts for both gain and internal loss \cite{CollettGardiner_1984, CollettGardiner_1985, Chelkowski_2007}.
From equations \eqref{eq_OPO_Vp}, \eqref{eq:internal_loss_def}, and \eqref{eq:opo_V_m} we find that the internal loss of a signal resonant squeezer is
\begin{align}
\Delta V_\text{internal loss, signal res} &> 1-\rho  
= \frac{\alpha}{g_{\mathrm{th}}}, \ \ \ &\ \frac{g}{g_{\mathrm{th}}} < 1 ,
\end{align}
where the low loss and near-threshold approximations are typical for signal-resonant squeezers, since the squeezing is maximized in these regimes. 
Since $\frac{g}{g_{\mathrm{th}}} <1$ below threshold, the internal loss of a signal-resonant squeezer is still bounded by the inverse of the nonlinearity to loss ratio $\Delta V_\text{internal loss, signal res} > \frac{\alpha}{g}$, same as the nonresonant squeezer. 
Intuitively, this implies that the signal-resonant squeezer provides an effective length enhancement but crucially \textit{does not change the maximum squeezing possible given the nonlinearity to loss ratio}. 
Although the above analysis is only valid in the high-finesse limit, numerical simulations show that even low-finesse signal resonances generate squeezing bounded by the nonlinearity to loss ratio. Furthermore, as the finesse is lowered (i.e. the escape efficiency is increased allowing larger round-trip gains while still operating sub-threshold), the resonant enhancement disappears and the low-finesse squeezer reduces to the nonresonant squeezer discussed earlier.

%%%%%%%%%%%%%%%%%%%%%%%%%%%%%%%%%%%%%%%%%%%%%%%%%%%%%%%
\subsubsection*{Pump Saturation} \label{Sec:Saturation}
%%%%%%%%%%%%%%%%%%%%%%%%%%%%%%%%%%%%%%%%%%%%%%%%%%%%%%%

\begin{figure}
    \centering
    \includegraphics[width=0.85\linewidth]{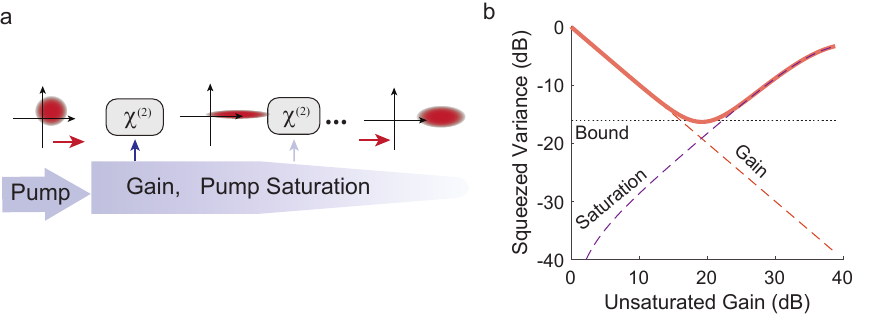}
    \caption{
    \label{fig:saturation} \textbf{Squeezer Saturation.} A coherent seed at the same frequency as the signal is input to the squeezer and gets amplified and depletes the pump’s energy. The pump’s noise gets transferred to the signal, degrading the squeezing. (b) Saturation limit on nonresonant squeezing, for a fractional leakage of $\frac{P_1^+(0)}{P_3(0)} = -35$ dB in a nonresonant squeezer with a shot-noise-limited pump beam, as a function of unsaturated gain $ G = e^{2gL}$.  Solid line: exact squeezing produced. Dotted line: bound on squeezing due to saturation (equation \eqref{eq:sat_bound}). Dashed lines show the simplified squeezing model of equation \eqref{eq:sat_deriv_start}, with terms of gain (orange) and saturation (purple). 
    }
\end{figure}

An amplifier’s output signal energy is limited by its pump energy - a squeezer \textit{saturates} if the amplified signal grows to levels comparable to the pump, which begins to deplete. Although the reduced pump power reduces gain, that is not the limiting factor on squeezing. In this section we show that, before significant pump depletion occurs, the quantum fluctuations of the pump and signal couple in a way that limits squeezing as illustrated in Fig.~\ref{fig:saturation}a.

Saturation can occur if a coherent seed is injected into the squeezer and amplified to bright levels. Coherent seeds are sometimes injected on purpose for applications that require bright squeezed light \cite{Bowen2021} or for phase-locking schemes \cite{Ozeki2022}, but can also be accidentally injected by leakage/backscatter in applications with squeezed vacuum. For example, McKenzie et al. \cite{Lam2004} observed noise degradation from local oscillator light backscattered off the photodetector and sent back into the squeezer, Chua et al. \cite{Chua2014} measured and quantified the effects of spurious light backscattered from the squeezer, and Stokowski et al. \cite{Stokowski2023} reported reduced performance due to coherent leakage into the squeezer from the pump generation process. Numerous other experiments have observed noise degradation due to pump depletion \cite{Peng2018, Chekhova2020}.

Suppose a degenerate seed is injected into the squeezer with power much less than the pump power $P_{1}(0) \ll P_3$ and phase relative to the pump $\theta_1$ - the squeezer deamplifies the power in one quadrature and amplifies the other 
\begin{align} \label{eq:sat_mean_evo}
    P_1^- &=  \frac{1}{G}P_1(0)\cos^2\theta_1\\
    P_1^+ &= {G}P_1(0)\sin^2\theta_1.
\end{align} 
As the amplified quadrature grows in power, noise from the pump transfers to the deamplified (squeezed) quadrature.
We will show in this section that the saturation contribution to noise is 
\begin{equation} \label{eq:sat}
\boxed{ \Delta V_\text{pump saturation} \gtrsim \frac{P_1^+}{2P_3(0)}  V_{3}^-},
\end{equation}
where $\frac{P_1^+}{P_3(0)}$ is the fraction of total power transferred from the pump to the signal.
and $V_3^-$ is the initial variance in the pump phase quadrature. 
As squeezers are pushed to higher gains, minimizing the seed power amplified within them becomes more important. This can be done by minimizing the injected seed power, such as with filters or anti-reflection designs, or by controlling the phase of the seed such that it does not get amplified to large levels. 

%%%%%%%%%%%%%%%%%%%%%%%%%%%%%%%%%%%
\paragraph{Nonresonant} \label{sec:Saturation_nonres}

The mean-field and quadrature variance evolution for nonresonant degenerate $\chi^{(2)}$ OPA conversion are derived in \cite{LiKumar1995} and simplified in the supplementary information% \ref{app:pump_depl_nonres} 
for the typical case of small starting seed power.
The squeezed variance evolves as
\begin{align} \label{eq:sat_deriv_start}
V_1^- &\approx \frac{P_1^+(0)}{P_1^+}V_1^-(0) + \bigg(\frac{P^+_1}{2 P_3(0)} + 2(gL)^2\frac{P^-_1}{ P_3(0)} \bigg)V_3^-, \qquad \substack{P_1^+(0) \ll P_1^+ < P_3(0) \\P_1^- \ll P_1^-(0) < P_3(0)} \\
&\approx \frac{1}{G}\bigg(V_1^-(0) \bigg) + G\bigg(\frac{P_1^+(0)}{2 P_3(0)}V_3^- \bigg) +  \frac{1}{G}\bigg(2(gL)^2\frac{ P_1^-(0)}{P_3(0)} V_3^-\bigg)   
\end{align}
The first term in equation \eqref{eq:sat_deriv_start} captures the expected squeezing gain of the "Gain" section %\ref{sec:gain}
while the second term represents the noise transfer due to seed amplification (and pump depletion) and the third term is the noise transfer due to seed deamplification. 
The amplification and deamplification noise contributions of equation \eqref{eq:sat_deriv_start} combine to the general bound of equation \eqref{eq:sat} in the limit of moderate nonlinearity ($2gL \sim 1$), and the amplification contribution tends to dominate since it scales with gain rather than the inverse of gain.
The evolution of the amplification contribution is shown in Figure~\ref{fig:saturation}b for coherent-state inputs for seed and pump $V_1^-(0) = 1, \  V_3^- = 1$. Initially, the variance decreases exponentially with gain. Eventually, however, the variance increases exponentially due to noise coupling caused by pump depletion - this is the noise transfer regime. At the transition between the two regimes, the minimum variance is achieved, giving the noise bound and optimal gain:
\begin{align} \label{eq:sat_bound}
V_\text{gain} + \Delta V_\text{pump saturation}& \ge 
 \sqrt\frac{2P_1^+(0)V_3^-}{P_3(0)} ,\\
G_{\mathrm{opt}} &= \sqrt{\frac{2P_3(0)}{P_1^+(0)V_3^-}}.
\end{align}

%%%%%%%%%%%%%%%%%%%%%%%%%%%%%%%%%%%%
\paragraph{Signal Resonant}

A coherent seed injected into a signal-resonant squeezer can build to high intracavity powers due to the combination of resonant feedback and gain, a process known as regenerative amplification \cite{siegman86}. Such coherent seeds have theoretically \cite{Bowen2006} and experimentally \cite{Lam2004, Bowen2008} been shown to introduce pump noise correlations in signal-resonant squeezers, similar to the nonresonant case.
In the supplementary information %\ref{app:signal-res_sat} 
we derive lower and upper bounds that constrain the saturation contribution to within a factor of 2. The lower bound is
\begin{align}\label{eq:upper_bound_opo_sat}
    \Delta V_\text{pump saturation} > \frac{4}{(1 + {\frac{g}{g_{th}}})^2} \frac{ P_1 }{ 2 P_{3,th}} V_{3}^- ,  \\
\end{align}
similar to equation \eqref{eq:sat} except with the pump power replaced by the threshold pump power $P_{3,th}$ and with an additional factor that reduces from 4 to 1 as threshold is approached.

%%%%%%%%%%%%%%%%%%%%%%%%%%%%%%%%%%%%%
\paragraph{Saturation without Seed}

It is possible for extremely high-gain squeezers to reach saturation even without a coherent seed - the squeezed vacuum itself gets amplified to bright levels in a regime known as optical parametric generation (OPG). 
The photon number and distribution effects for OPG were measured experimentally in a nonresonant system \cite{Chekhova2020} and calculated in \cite{Xing2023, Yanagimoto2022}. Correlations between pump and signal were calculated for a fully-resonant system in \cite{Bowen2006}. 
We do not go into detail on these noise effects where the linearized quantum treatment breaks down. 
However, the total squeezed power $P_{SV}$ cannot exceed the pump power $P_3$ by energy conservation, yielding a conservative squeezing bound 
\begin{align} \label{eq:P_SV}
    P_3 &\gg P_\text{SV}, \\
    &\gg \frac{G-\frac{1}{G}}{4}h\nu\Delta\nu,
\end{align}
where the second line expresses the power of the squeezed vacuum in terms of the squeezing bandwidth $\Delta \nu$, photon energy $h\nu$, and gain $G$.  
Rearranging equation \eqref{eq:P_SV} gives the conservative squeezing bound:  
\begin{equation}
\Delta V_\text{pump saturation} \gg \frac{ h\nu\Delta\nu}{4P_3}.
\end{equation}
In practice, extremely high gains are required to reach this squeezing bound and it is unlikely to limit practical integrated squeezers.

%%%%%%%%%%%%%%%%%%%%%%%%%%%%%%%%%%%%%%%%%%%%%%%%%%%%%%%
\subsubsection*{Phase Noise} \label{sec:phasenoise}
%%%%%%%%%%%%%%%%%%%%%%%%%%%%%%%%%%%%%%%%%%%%%%%%%%%%%%%

Applications and detection schemes for squeezed light must be phase sensitive in order to fully utilize the squeezed quadrature. Phase noise, or fluctuations in the relative phase between the squeezed quadrature and the detection quadrature, degrades squeezing by mixing in part of the antisqueezed quadrature’s noise. Phase noise scales with gain and is therefore closely linked to the generation of squeezed light.

\begin{figure}
    \centering
    \includegraphics[width=0.65\linewidth]{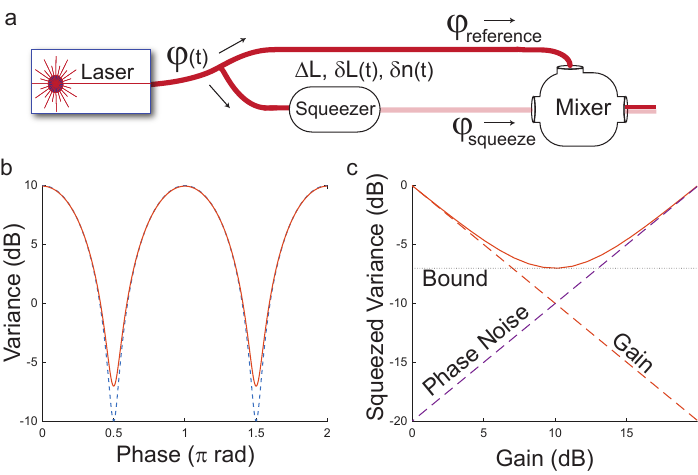}
    \caption{    \label{fig:phasenoiseconcept} \textbf{Phase error.}
        (a) Contributions to phase error between the reference beam and squeezed beam - laser phase noise and path length difference, path length fluctuations, and refractive index fluctuations. (b) Quadrature variance vs quadrature angle (equation \ref{eq:phasenoise} for different quadrature angles), assuming a gain $G=10$ and phase noise $\delta\theta = 100$ mrad. (c) Squeezing as a function of gain, assuming fixed phase noise $\Delta\theta = 100$ mrad. The horizontal dotted line shows the bound (equation \eqref{eq:phasenoise_bound}) on squeezing given the phase noise, and the dashed lines show the contributions from phase noise (equation \eqref{eq:phasenoise_approx}) and gain (equation \eqref{eq:V_gain}). 
}
\end{figure}

The standard deviation in phase error $\delta\theta = \sqrt{\langle (\phi_\text{squeeze} - \phi_\text{reference})^2 \rangle}$ between the squeezed and reference beams with phases $\phi_\text{squeeze}, \phi_\text{reference}$ degrades measured squeezing according to \cite{Zhang_2003_phasenoise, Furusawa2006}
\begin{align} \label{eq:phasenoise}
    V &= \cos^2(\delta\theta) V^- + \sin^2(\delta\theta) V^+ \\
    &\gtrsim V^- + \frac{1}{V^-} \sin^2(\delta\theta), \qquad \qquad \delta\theta \ll 1, \ \ V^-V^+ \geq 1 \nonumber
\end{align}
and is plotted in Fig.~\ref{fig:phasenoiseconcept}b,c. The first term is the ideal squeezing without any phase error, and the second term captures how phase error mixes in antisqueezing which, by the Heisenberg uncertainty principle, is at least as large as the inverse of the squeezing as established in the "Gain" section.% \ref{sec:gain}. 
Therefore the phase noise contribution is
\begin{equation} \label{eq:phasenoise_approx}
\boxed{ \Delta V_\text{phase noise} = G\sin^2\delta\theta }.
\end{equation}
The transition between the squeezing and phase noise regimes, depicted in Figure ~\ref{fig:phasenoiseconcept}c yields the minimum variance and optimal gain
\begin{align} \label{eq:phasenoise_bound}
V_\text{gain}  + \Delta V_\text{phase noise} \ge \sin 2\delta\theta, \\
G_{\mathrm{opt}} = \cot\delta\theta.
\end{align}

%%%%%%%%%%%%%%%%%%%%%%%%%%%%%%%%%%%
\paragraph{Sources of Phase Noise}

Since the phase of a typical laser drifts at a rate proportional to its linewidth, a single narrow-linewidth laser is typically used to create both the squeezed and reference beams. This way, the two beams share the same global phase drift and therefore have minimal relative phase drift. However, path-length mismatch between the two beam paths converts laser frequency noise into phase noise, and phase fluctuations due to path-length fluctuations, refractive-index fluctuations, and even phase modulations and noise in the locking process can all introduce relative phase error \cite{Sigg2013}.
Resonant cavities can further increase the effects of phase noise\cite{Mavalvala2014}, since the larger effective lengths of cavities increases sensitivity to path length mismatch, the multiple round trips enhances thermorefractive noise, and the need for a separate cavity phase lock implies additional modulation that can contribute to phase error.

% \textbf{Laser Phase Noise and Path Mismatch}
In the typical case where both the squeezer pump and the reference beam are derived from the same laser, path differences between the two beams can introduce laser phase noise. The short length-scales inherent to nanophotonic chips greatly reduce phase noise induced by path-length mismatch compared to bulk systems. 
Assuming a laser with full-width at half-maximum linewidth $\Delta\nu$ and white frequency-noise spectrum , the standard deviation phase error $\theta$ between two unequal paths with difference $\Delta L$ in material with group velocity $v_g$ is \cite{Domenico2010}
\begin{equation}
\delta\theta_\text{laser} = \sqrt{2\pi \Delta\nu \frac{|\Delta L|}{v_g}}.
\end{equation}
For a low-noise laser with linewidth $\Delta\nu = 1$ kHz and a chip-scale path length mismatch $\Delta L = 1$ mm in material with group velocity $v_g = 0.5c$, the phase error is $\Delta\theta = 0.2$ mrad. This small phase error would limit squeezed variance to $V_{min} = -34$  dB.
Most lasers do not have flat frequency-noise spectra and instead have more noise at low frequencies. Low-frequency noise can be suppressed by narrow-linewidth cavity filters or active phase locks. 

% \textbf{Path Noise, Thermorefractive Noise, and Modulator Noise}
Relative phase fluctuations between the reference and signal can also arise from path and refractive index fluctuations. For example, if the relative path length fluctuates with standard deviation $\delta L$ 
then the relative phase fluctuates by
\begin{equation}
    \delta\theta_\text{path} = \frac{2\pi n}{\lambda} \delta L ,
\end{equation}
where $n$ is the effective refractive index and $\lambda$ is the wavelength of light.
Although large free-space squeezers have observed noise due to mechanical vibrations and drift \cite{Chua2014, Sigg2013}, the smaller and more stable length scales of nanophotonic chips largely suppress path noise from mechanical fluctuations and vibrations. However, effective path fluctuations can also arise from thermal noise, thermal crosstalk, and electro-optical sources.
% \textbf{Thermal Noise}
Thermal noise is a source of noise present in on-chip photonics that is negligible in free space. It arises due to microscopic thermodynamic fluctuations in temperature, and introduces phase fluctuations on optical fields. The phase fluctuations are caused by the temperature-dependence of the refractive index by material polarizability, stress, and, in the case of electro-optic materials, charge distributions \cite{Kippenberg2025, yin2025fundamentalphasenoisefilm}.
Although thermal noise is often negligible for frequencies beyond one MHz, it can contribute to low frequency phase noise, particularly in resonant systems. 
The phase modulators employed in chip-scale squeezers can introduce additional fluctuations. Electro-optic modulators, if not properly grounded, can pickup stray microwave signals in the environment and modulate them onto the light. Thermo-optic modulators are prone to thermal crosstalk, where thermal modulations on one part of the chip unintentionally imprint modulations onto other beam paths.

% \textbf{Phase Locking Error}
Squeezed light systems typically require a phase lock between the squeezed beam and the reference beam to align and maintain the relative phase in the presence of phase noise. 
The interference of the two beams generates an error signal that is then fed back into a phase shifter to correct for any phase drift. Phase locking effectively suppresses phase noise within the lock bandwidth, but imperfect phase detection adds noise to the feedback. 
For example, in the supplementary information %\ref{app:phaselock} 
we show that a phase lock with open-loop (no feedback) signal-to-noise ratio $\text{SNR}_0$ and feedback-loop gain $L_0$ contributes phase error
\begin{align}
    \delta\theta_\text{lock SNR} = \delta\theta_{in} \frac{\sqrt{1 + \frac{L_0^2}{SNR_0}}}{(1+L_0)},
\end{align}
where $\delta\theta_\text{in} = \sqrt{ \delta\theta_\text{laser}^2 + \delta\theta_\text{path}^2}$ is the phase noise prior to the lock. 
If the phase detection is noisy (small $\text{SNR}_0$), then some of that noise is fed back onto the light, and, for large feedback-loop gain, directly limits phase error $\delta \theta _\text{lock SNR} \approx \frac{\delta\theta_\text{in}}{\sqrt{SNR_0}}$. 
% Now introduce modulation
The lock quadrature is determined by the details of the error signal. Error signals generated by a DC signal (or intensity modulation) can lock in-between the bright/dark fringe of the signal-reference interference, while those generated by a relative-phase modulation can lock to the bright/dark fringe, and a combination of the two can lock to any quadrature \cite{Chelkowski_2007}. 
However, there is a tradeoff between modulation depth and signal-to-noise-ratio (SNR) in phase-modulated locks - the modulation depth must be kept small in order to not generate phase error by itself, while at the same time sufficiently large to measure phase drift. 
For example, a lock with phase modulation at frequency $\Omega$ outside the locking bandwidth and modulation depth $m$ induces root-mean-square phase error and open-loop SNR
\begin{align}
\delta\theta_\text{lock mod} &= \sqrt{ \langle |m\sin\Omega t|^2 \rangle} = \frac{m}{\sqrt 2},\\
SNR_0 &\propto m^2.
\end{align}
Although it is possible to generate an error signal for locking with the anti-squeezing itself \cite{Lam2004}, this typically requires high modulation depth to achieve sufficient SNR for locking, which introduces phase error. Instead, squeezed systems often allow a small amount of optical power into the squeezer, thereby generating a displaced squeezed state where the displacement and squeezing phases are fixed relative to each other. Then, locking to the squeezed noise is as straightforward as locking the reference beam to the displacement of the squeezed beam. 
Alternative locking schemes have been demonstrated that, in ideal operation, generate a locking error signal without adding power or noise to the squeezed light field\cite{Bowen2002, hirota2025}.

%%%%%%%%%%%%%%%%%%%%%%%%%%%%%%%%%%%%%%%%%%%%%%%%%%%%%%%
\subsubsection*{Summary of Squeezed Light Generation}
%%%%%%%%%%%%%%%%%%%%%%%%%%%%%%%%%%%%%%%%%%%%%%%%%%%%%%%

So far, we’ve discussed five challenges that must be overcome in order to generate highly squeezed light. We noted that the squeezing cannot exceed the gain, though use of resonances and high pump powers can increase gain exponentially. Even in the presence of high gain, the squeezing will be limited by internal loss, related to the nonlinearity to loss ratio of the squeezer. If any coherent seed is input to the squeezer, such as for use in phase locking or injected by stray reflections, then any amplification of the seed will also transfer pump noise. Inefficiency in outcoupling the squeezed light contributes the well-known vacuum noise we referred to as external loss. Lastly, phase error couples noise of the antisqueezed quadrature and scales with gain.

\begin{figure}
    \centering
    \hspace*{-0.05\linewidth}
    \includegraphics[width=1.1\linewidth]{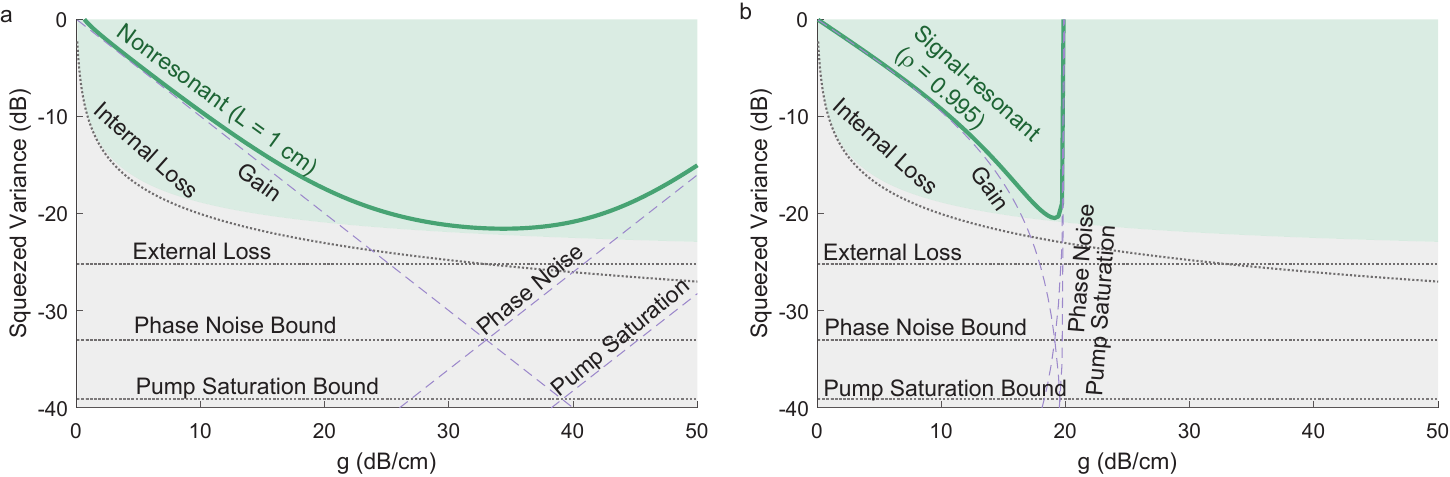}
    \caption{\label{fig:sqz_gen} \textbf{Fundamental bounds on squeezed light generation.} Plotted as a function of gain rate g, for different kinds of squeezers. Light green shaded region represents the feasible squeezing region, for all possible squeezer designs (nonlinear length, resonance, etc). Large dotted lines show noise bounds, for all possible squeezer designs. Solid green and faint dotted lines show the total expected squeezing and the gain-dependent noise limitations for specific squeezer designs. Plotted for internal loss rate $\alpha = 0.1$ dB/cm, external loss $1-\eta_e = -25$ dB, phase noise $\delta\theta = 0.5$ mrad, signal leakage into squeezer $\frac{P_1^+(0)}{P_3(0)} = -75$ dB. (a) Nonresonant squeezer with nonlinear length $L = 1$ cm. (b) Signal-resonant squeezer with escape efficiency $\rho = 0.995$.
        }
\end{figure}

Fig.~\ref{fig:sqz_gen} plots each of these noise contributions, for optimistic yet feasible experimental numbers, for (a) a nonresonant OPA squeezer with $L = 10$ mm nonlinear length and (b) a signal-resonant squeezer with escape efficiency $\rho = 0.995$. 
The light green feasible squeezing region is identical in both squeezer designs. In other words, given the effective internal and external loss as well as the phase error and saturation bounds, the performance of \textit{any} squeezer design, regardless of nonlinear length or resonance conditions, is restricted to be within the feasible region. The actual generated squeezing, shown by the solid green lines, differs in the two cases. 
Consider Fig.~\ref{fig:sqz_gen} (a). For low gain rates, the squeezing is limited by the gain and variance decreases exponentially with gain rate. For moderate gain rates, the squeezing becomes limited by a combination of internal and external loss and becomes constant. For very high gain rates, the squeezing degrades due to phase noise. 
The noise limits for a signal-resonant squeezer are very similar to those shown in Fig.~\ref{fig:sqz_gen}(a), with the slight difference that the gain (and phase noise, saturation) diverge as the threshold gain rate is approached.
Note that a pump resonant squeezer would perform similarly to the nonresonant squeezer in Fig.~\ref{fig:sqz_gen} (a), with a rescaling of the X-axis to be the effective gain rate.

%%%%%%%%%%%%%%%%%%%%%%%%%%%%%%%%%%%%%%%%%%%%%%%%%%%%%%%%%%%%%%%%%%%%%%%%%%%%%%%%%
\subsection*{Squeezed Light Detection}
%%%%%%%%%%%%%%%%%%%%%%%%%%%%%%%%%%%%%%%%%%%%%%%%%%%%%%%%%%%%%%%%%%%%%%%%%%%%%%%%%

The goal of squeezed light detection is to amplify the squeezed signal by mixing it with a reference beam, and then detect the squeezed noise without also detecting much electronic noise from the electronics or intensity noise from the reference beam. In the following sections, we examine the noise contributions of two common squeezed light detection methods: balanced homodyne detection and (degenerate) parametric homodyne detection. The main noise contributions in both cases are due to the intensity noise of the reference beam, detection loss, and electronic noise.

Balanced homodyne detection is the most common detection method for measuring squeezed light. In a balanced homodyne detector, the squeezed light is mixed with a bright reference beam known as the local oscillator on a 50-50 beamsplitter, and the two resulting outputs are detected and their photocurrents subtracted by a balanced detector \cite{YuenShapiro1979}. The relative phase between the local oscillator and the squeezed light determines what quadrature appears on the resulting photocurrent signal. A key feature of balanced homodyne detection is that, in ideal performance, it is immune to the intensity noise of the local oscillator \cite{Loudon1986}.

In contrast to conventional homodyne detection, which uses linear mixing with a strong reference beam to boost the quadrature of interest to detectable levels, parametric homodyne detection uses nonlinear mixing with a strong reference beam to amplify the quadrature of interest to bright levels \cite{Pe'er2018, Furusawa2020, Nehra2022}. Specifically, the pump of a high-gain optical parametric amplifier serves as the reference beam, and its phase determines the measurement quadrature. This is essentially a squeezer, except the phase is chosen not to deamplify (squeeze) the noise but rather to amplify both the squeezed signal and its noise equally, generating a readily-measured bright state with the same signal-to-noise ratio as the squeezed signal. A key feature of parametric homodyne detection is that, after amplification, the signal is tolerant to detection loss and other noise sources.

\begin{figure}
    \centering    \includegraphics[width=0.75\linewidth]{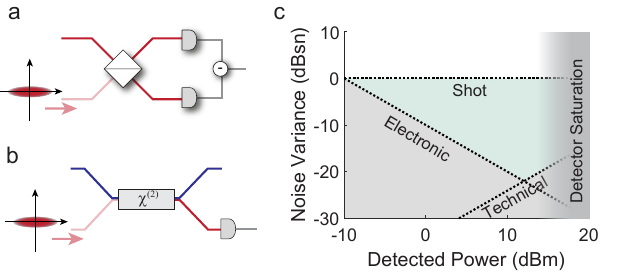}
    \caption{\label{fig:detnoise} 
    \textbf{Squeezed light detection}.
     (a) Balanced homodyne detection mixes the squeezed signal (light red) with a strong local oscillator reference beam (red) and detects the difference photocurrent out of two photodetectors. 
    (b) Parametric homodyne detection mixes the squeezed signal (light red) with a strong pump (blue) and detects the amplified signal.
    (c) Contributions to detection noise, in decibel shot-noise units, as a function of detected power. Parameters assumed include electronic noise with 20 dB shot noise clearance at 10 mW, excess relative intensity noise of -150 dBc/Hz for $\lambda = 1550$ nm reference beam, suppressed by balanced detection with a common mode rejection ratio of 40 dB, and detector saturation power roughly 15 dBm.
    }
\end{figure}

It is even possible to combine some of the desirable features of each detection scheme by employing an optical preamplifier prior to balanced homodyne detection \cite{Chekhova2021, Furusawa2025}. Unlike in parametric homodyne detection, where the amplifier both selects the measurement quadrature and amplifies it to macroscopic levels, this optically preamplified balanced homodyne detection only uses a relatively low-gain amplifier to achieve partial noise tolerance and the measurement quadrature is still selected by the local oscillator. Although here we do not explicitly consider this and other detection schemes, the noise contributions we discuss are readily generalizable (see Methods section "Applying and Visualizing the Model").% \ref{sec:App:preampBHD}).

Figure ~\ref{fig:detnoise}c depicts the scaling of different detection noise contributions as a function of detected optical power, for some reasonable device parameters. Relative to shot noise, as detected power increases, electronic noise becomes less significant and technical noise more significant. Intuitively, these relationships arise from the contributions to power spectral density (PSD) - electronic noise PSD doesn't depend on optical power at all, shot noise PSD scales linearly with average power, and excess classical noise PSD, which is proportional to the mean square of classical power fluctuations, scales quadratically with power. Past some power, the detector saturates. The shaded green region represents the range of noise variances where it is possible to measure the reduced shot noise of squeezed light.

%%%%%%%%%%%%%%%%%%%%%%%%%%%%%%%%%%%%%%%%%%%%%%%%%%%%%%%
\subsubsection*{Intensity Noise}
%%%%%%%%%%%%%%%%%%%%%%%%%%%%%%%%%%%%%%%%%%%%%%%%%%%%%%%

The intensity noise of the reference beam can contribute to measurement noise in two ways. First, the reference intensity noise creeps in if the detection is not perfectly balanced. Second, even if the detection is balanced, the intensity noise of the reference is detected if the signal power is substantial compared to the reference power. We now briefly describe intensity noise, before quantifying these two noise contributions for balanced detection and parametric homodyne detection.

% %%%%%%%%%%%%%%%%%%%%%%%%%%%%%%%
% \subsubsection*{Intensity Noise}

The intensity noise of laser light comprises of both the quantum fluctuations of shot noise and the excess classical fluctuations of technical noise, which often arise from the laser itself. Relative to shot noise, the intensity noise is
\begin{equation}
V^X=  V^X_\text{shot noise} + V^X_\text{classical} =1 +  \frac{P}{P_{\text{RIN}_\text{cl=sn}}}
\end{equation}
where $P_{\text{RIN}_\text{cl=sn}} = \frac{2h\nu}{\text{RIN}_\text{cl}}$ is the power at which the shot noise and classical noise are equal\cite{Coldren_RIN}, $\text{RIN}_\text{cl}$ is the excess relative intensity noise of the light and $h\nu$ is the photon energy. 
Because the ratio of classical intensity noise to shot noise scales with detected power, it can become the dominant noise source at high optical powers. 
The excess classical noise can be reduced with a low-noise laser, balanced detection, narrowband filtering, or a noise-eater\cite{Bachor2019}, and can be avoided by shifting the signal of interest to a higher frequency where technical noise tends to be lower \cite{Hobbs1997}. Only the intensity noise within the measurement bandwidth degrades measured squeezing, in contrast with the laser phase noise discussed in the "Phase Noise" Section.% \ref{sec:phasenoise}.

%%%%%%%%%%%%%%%%%%%%%%%%%%%%%%%
\paragraph{Balanced Homodyne Detection}
% \textbf{Imbalance:}
In balanced detection, each detector records the local oscillator's intensity noise and the resulting difference photocurrent cancels the two noise contributions - yielding common-mode rejection ratios in excess of 50 dB \cite{Stefszky_2012}. 
However, any imbalance, due to imperfect mixing of beams on the beamsplitter or non-identical photodiodes, leads to residual intensity noise in the difference photocurrent. 
For example, for beamsplitter splitting ratio is $\eta_\text{bs}$, detector responsivities $ R_i$, and local oscillator power $P_\text{LO}$, the mean difference photocurrent is 
\begin{align}
    i_\text{BHD} = [R_1 \eta_\text{bs} - R_2 (1-\eta_\text{bs})]P_\text{LO}
\end{align} 
and the corresponding intensity noise contribution is
\begin{equation}
    \Delta V_\text{imbalance} = V^X_{\mathrm{LO}} (1-\mathcal{V}^2_\text{imbalance})
\end{equation}
where $V^X_{\mathrm{LO}}$ is the intensity noise (shot-noise and classical) of the local oscillator and the imbalance term (also known as common-mode rejection ratio) is $1-\mathcal{V}_\text{imbalance}^2 = \frac{R_1 \eta_\text{bs} -  R_2 (1-\eta_\text{bs})}{R_1 \eta_\text{bs} + R_2 (1-\eta_\text{bs})}$.
In contrast with free space beamsplitters, which can typically be angle-tuned to achieve perfect 50-50 splitting, integrated photonic beamsplitters tend to have fixed splitting ratios after fabrication. This poses a problem for squeezed light systems with fabrication variances that exceed splitting ratio tolerances. One workaround is to employ Mach-Zehnder interferometers as effective tunable beamsplitters that can be controlled to compensate for fabrication imperfections \cite{Miller2015, Stokowski2023, Qi:25}. Another workaround is to adjust the relative gain on the two photocurrents before differencing to compensate for imbalance. 
 Note that the imbalance may be wavelength-dependent - broadband squeezed vacuum or laser amplified spontaneous emission may exceed the 50-50 performance bandwidth of integrated splitters and therefore directly introduce intensity noise or lead to improper balancing. An optical bandpass filter can effectively suppress this broadband noise.

%%%%%%%%%%%%%%%%%%%%%%%%%%%%%%%
% \textbf{Measurement Saturation: Reference Beam Noise}
Balanced homodyne detection relies on the local oscillator being much brighter than the squeezed beam. If the squeezed beam has substantial power, then it can introduce intensity noise from both itself and from the local oscillator.
The measured variance of a balanced homodyne detector with squeezed beam power $P_1 = P_1^- + P_1^+$ and local oscillator power $P_{\mathrm{LO}}$ and variances $V_{\mathrm{LO}}^X, V_{\mathrm{LO}}^Y$ is given by
\begin{equation}
V_\text{BHD} = V^- + \frac{ P_1^- }{P_{\mathrm{LO}}}V^X_{\mathrm{LO}} + \frac{P_1^+}{P_{\mathrm{LO}}} V^Y_{\mathrm{LO}}.
\end{equation}
The first term is the desired squeezing term and the last two terms contain the shot noise of the local oscillator mixed in by the power in the squeezed beam. The larger the signal power in one quadrature, the more the noise of the corresponding local oscillator quadrature is detected. For a shot-noise limited local oscillator, the noise contribution reduces to simply the ratio of signal power to local oscillator power. 
The total local oscillator noise contribution in balanced homodyne detection is nothing but the combination of intensity noise due to imbalance and noise due to signal power:
\begin{equation}
\boxed{
\Delta V_\text{intensity noise, BHD} = V^X_{\mathrm{LO}} (1-\mathcal{V}_\text{imbalance}^2) + 
 \frac{ P_1^- }{P_{\mathrm{LO}}}V^X_{\mathrm{LO}} + \frac{P_1^+}{P_{\mathrm{LO}}} V^Y_{\mathrm{LO}}
}
\end{equation}

%%%%%%%%%%%%%%%%%%%%%%%%%%%%%%%
\paragraph{Parametric Homodyne Detection} \label{sec:PHD_sat}

In parametric homodyne detection, the power-dependent gain of the amplifier converts pump intensity fluctuations into gain fluctuations that cause intensity noise in the amplified signal. 
The output intensity noise from a phase-sensitive amplifier with pump-power-dependent gain $G_\mathrm{det}(P)$ and pump intensity variance $V_3^X(0)$ is
\begin{align} 
        V_\text{amplified} &\approx  G_{\mathrm{det}}V_\text{1,det}^+(0) +  2\bigg(\frac{\partial G_{\mathrm{det}}}{\partial P}|_{P_\text{3, det}}\bigg)^2  P_\text{1,det}^+(0) P_\text{3,det}  V_\text{3,det}^X,
\end{align}
where the first term is the desired amplified squeezed signal and the second term is the pump intensity noise contribution, for a signal with initial power $P_\text{1,det}^+(0)$ in the amplification quadrature. Note that the amplification quadrature for parametric homodyne detection of a squeezed state is the squeezed quadrature from before $P_\text{1,det}^+(0) = P_1^-$, if the phase is properly set to measure squeezing. The detection gain in PHD should be sufficiently large such that most of the detected power comes from just the amplified quadrature.
With respect to the shot noise level of the squeezed signal (prior to PHD), the noise contribution from pump intensity noise is
\begin{align} \label{eq:V_PHD}
    \boxed{
    \Delta V_\text{intensity noise, PHD} \approx   2\bigg(\frac{\partial G_{\mathrm{det}}}{\partial P}|_{P_\text{3, det}}\bigg)^2  \frac{P_1^-}{ G_\text{det} } P_\text{3, det} V_\text{3,det}^X
    }.
\end{align}
For example, evaluating equation \eqref{eq:V_PHD} for the exponential gain scaling of a nonresonant squeezer yields
\begin{align} \label{eq:V_PHD_nonres}
    V &\approx \frac{1}{2}(\ln G_{\mathrm{det}})^2\frac{P_1^-}{P_\text{3, det}}V_\text{3,det}^X 
\end{align}
Similar to balanced homodyne, the reference beam’s intensity noise couples in scaled by the ratio of signal to reference power. 
In principle, detecting the pump's intensity fluctuations and subtracting their effect on the signal could eliminate this noise\cite{Bowen2006}.
More details and extension to resonant amplifier designs are in the supplementary information.% \ref{app:PHD intensity noise}.

%%%%%%%%%%%%%%%%%%%%%%%%%%%%%%%%%%%%%%%%%%%%%%%%%%%%%%%
\subsubsection*{Detection Loss and Mode Mismatch} \label{Sec:DetectionLoss}
%%%%%%%%%%%%%%%%%%%%%%%%%%%%%%%%%%%%%%%%%%%%%%%%%%%%%%%

Detection loss degrades squeezing the same as external loss: by introducing vacuum noise by the beamsplitter expression of Equation  \eqref{eq:external_loss}. Mode mismatch between the reference beam and squeezed beam, coupling loss and reflections off the photodetectors, and imperfect quantum efficiency of the photodetectors are common sources of detection loss. 

% \textbf{Mode Mismatch}

Mode matching, or the spatial and temporal overlap of the reference beam with the squeezed beam, is essential for detection. The degree of mode matching is characterized by the normalized interference fringe contrast known as visibility $\mathcal{V}$ and the degradation of measured squeezing is simply \cite{Bachor2019}
\begin{equation}
V = \mathcal V^2 V^- + (1-\mathcal V^2).
\end{equation}

In free space systems, mode matching can be difficult and alignment-sensitive. In contrast, integrated photonic waveguides can support only a few discrete modes and no alignment is necessary. Mode filters or polarization controllers can ensure the mode of the reference beam matches that of the squeezed light. In some integrated platforms, such as lithium niobate, birefringence in waveguide bends can induce problematic coupling between polarization modes.

% \textbf{Detector Quantum Efficiency}

Another key aspect of detection loss is detector quantum efficiency. The purpose of a photodetector is to convert photons into ejected electrons that form a measurable photocurrent. The quantum efficiency of the detector is the fraction of incident photons that excite  measurable electrons. Good photodetectors can have quantum efficiencies higher than 98\% \cite{Shen2022}. Reflections at the detector's interface, imperfect absorption of photons within it, and electron-hole recombination in the resulting photocurrent all degrade the quantum efficiency. 
Detector quantum efficiency $\eta_{QE}$ contributes vacuum noise the same way as other loss channels
\begin{equation}
V_{QE} = \eta_{QE} V^- + (1-\eta_{QE}) .
\end{equation}
The visibility and quantum efficiency can be combined to yield the total detection efficiency \cite{Bachor2019}:
\begin{equation}
\eta_{\mathrm{det}} = \eta_{QE}\mathcal{V}^2.
\end{equation}
and so the noise contribution of detection loss, similar to that from external loss, is
\begin{equation}
 \boxed{\Delta V_\text{detection loss, BHD} = 1-\eta_{\mathrm{det}} }.
\end{equation}

%%%%%%%%%%%%%%%%%%%%%%%%%%%%%%%
% \subsubsection*{PHD}

 The phase-sensitive amplification process boosts the initially fragile squeezed measurement signal and noise to macroscopic levels far above the vacuum noise level, so any noise introduced after amplification (such as by loss) does not significantly degrade the signal to noise ratio. Intuitively, parametric homodyne detection is sensitive to loss before amplification, partially sensitive to loss during amplification, and robust to loss after amplification. The total internal loss contribution to noise discussed here is closely related to the noise figure of the optical amplifier, which is how much it degrades the signal-to-noise ratio compared to an ideal, noiseless amplifier \cite{OPANFtheory}. 
The detection loss contribution of parametric homodyne detection, relative to the input shot noise, is 
\begin{equation}
\boxed{
\Delta V_\text{detection loss, PHD} = \frac{\alpha}{g_\text{det}-\alpha} + \frac{1-\eta_{\mathrm{det}}}{G_{\mathrm{det}}}
}.
\end{equation}
The first term captures the effect of internal loss during amplification, which scales as the inverse of the nonlinearity to loss ratio. This bound, and the intuition for it, follow analogously to the squeezer internal loss discussed in Section "Internal Loss" %\ref{Sec:internalloss}
and also applies to resonant amplifiers.
In practice, this internal loss bound of the detection amplifier shouldn’t substantially limit measured squeezing, since the squeezing amplifier is subject to the same internal loss bound.
After amplification, the relative noise contribution is reduced by the gain factor. Parametric homodyne detection is therefore particularly appealing for applications limited by detection loss.

%%%%%%%%%%%%%%%%%%%%%%%%%%%%%%%%%%%%%%%%%%%%%%%%%%%%%%%
\subsubsection*{Electronic Noise}
%%%%%%%%%%%%%%%%%%%%%%%%%%%%%%%%%%%%%%%%%%%%%%%%%%%%%%%

Electronic noise, arising from fluctuations in the detector circuitry, is the dominant noise source at low optical powers \cite{Schiller1998}. Relative to shot noise, electronic noise decreases in significance with detected power and can be expressed as 
\begin{align}\label{eq:electronic}
   \boxed{
   \Delta V_\text{electronic noise, BHD} = \frac{\text{NEP}^2 \eta_\text{QE} }{2 h \nu} \frac{1}{P_\text{det}}
   },
\end{align}
for detected power $P_\text{det}$ and detector quantum efficiency $\eta_\text{QE}$ and noise-equivalent power NEP.
There is typically excess electronic noise at low frequencies, which is avoided in measurements performed at higher frequencies. 
Parametric homodyne detection is tolerant to electronic noise the same way it is tolerant to detection loss - the relative contribution of electronic noise is divided by the gain of the parametric detection amplifier: 
\begin{equation}
\boxed{
 \Delta V_\text{electronic noise, PHD}
 = \frac{1}{G_{\mathrm{det}}}\frac{\text{NEP}^2 \eta_\text{QE} }{2 h \nu} \frac{1}{P_\text{det}} }.
\end{equation}

%%%%%%%%%%%%%%%%%%%%%%%%%%%%%%%%%%%%%%%%%%%%%%%%%%%%%%%
\subsubsection*{Summary of Squeezed Light Detection}
%%%%%%%%%%%%%%%%%%%%%%%%%%%%%%%%%%%%%%%%%%%%%%%%%%%%%%%

Numerous noise sources in the detection chain can obscure squeezing, albeit with slightly different effects depending on the detection scheme. Noise from the reference beam appears if the detection is not balanced or if the squeezed signal has significant power. Detection loss and mode mismatch contribute vacuum noise. Lastly, electronic noise forms a baseline noise level in any photodetector circuit. Although balanced homodyne detection is a well-established detection method for squeezed light, parametric homodyne detection’s tolerance to certain noise contributions makes it an attractive candidate as well.

%%%%%%%%%%%%%%%%%%%%%%%%%%%%%%%%%%%%%%%%%%%%%%%%%%%%%%%%%%%%%%%%%%%%%%%%%
\section*{Discussion}
%%%%%%%%%%%%%%%%%%%%%%%%%%%%%%%%%%%%%%%%%%%%%%%%%%%%%%%%%%%%%%%%%%%%%%%%%

We derived and summarized simple analytic expressions for many noise contributions to squeezed light generation and detection. Substituting them into our simple model, we arrive at
\begin{align} \label{eq:full_noise_model}
V &\approx \big(V_\text{gain} + \Delta V_\text{internal loss} + \Delta V_\text{pump saturation} + \Delta V_\text{external loss} + \Delta V_\text{phase noise}  \big)\\ &\ \ \ \ \  + \big(   \Delta V_\text{intensity noise} + \Delta V_\text{detection loss}   +\Delta V_\text{electronic noise}  \big) \nonumber \\
&\approx 
\frac{1}{G} + \frac{\alpha}{g} 
 + \frac{P_1}{2P_3}V_3^-  + (1-\eta_{e}) +  G\sin^2\delta\theta \\ & \ \ \ \ \ +
 \begin{cases}
  V^X_{\mathrm{LO}} (1-\mathcal{V}_\text{imbalance}^2) + 
 \frac{ P_1^- V^X_{\mathrm{LO}} + P_1^+ V^Y_{\mathrm{LO}} }{P_{\mathrm{LO}}}
 +  (1-\eta_{\mathrm{det}}) + \frac{\text{NEP}^2 \eta_\text{QE} }{2 h \nu} \frac{1}{P_\text{det}}, \ \ \ \ \ \ \text{BHD} \nonumber  \\
     2\bigg(\frac{\partial G_{\mathrm{det}}}{\partial P}|_{P_\text{3, det}}\bigg)^2  \frac{P_1^-}{ G_\text{det} } P_\text{3, det}  V_\text{3,det}^X 
 +  \frac{\alpha}{g_\text{det}} + \frac{1-\eta_{\mathrm{det}}}{G_{\mathrm{det}}}  + \frac{\frac{\text{NEP}^2 \eta_\text{QE} }{2 h \nu} \frac{1}{P_\text{det}} }{G_\text{det}}, \  \  \text{PHD} \nonumber
 \end{cases} 
\end{align}
and the more precise model valid for significant noise mixing terms can be obtained with equation \eqref{eq_noise_sum}.
We derived expressions for the various noise terms for the nonresonant, signal resonant, and pump resonant squeezers and parametric and balanced homodyne detectors. Table \ref{summary_table} summarizes the different noise terms and serves as a practical guide to squeezer design. The first four columns describe the noise model, the fifth column estimates squeezer performance bounds, and the last two columns provide guidance on mitigating the noise limitations.

For the most part, this model is general to \textit{any} nonlinear optical squeezer. 
In adding the different noise variances, the model assumes that each of the noise contributions are uncorrelated with all the other noise contributions - this is typically the case in practice.
% However, for example, cavities are capable of partially converting phase noise into intensity noise, which could correlate those two noises\cite{Bachor2019}.

\begin{landscape}
% \begin{center}
\begin{table} 
% \begin{sidewaystable} 
\footnotesize
\hspace*{-0.05\linewidth}
\begin{tabular}{||m{0.07\textwidth} m{0.27\textwidth} m{0.27\textwidth} m{0.32\textwidth} m{0.06\textwidth} m{0.12\textwidth} m{0.33\textwidth}||} 
 \hline
 Parameter & Description & Variables & Noise Contribution & Noise Bound & Optimal Parameter & Mitigation Strategy \\ [0.5ex] 
 \hline\hline
 Gain & Initial squeezing is limited to the inverse of the gain & Gain $G$, input variance $V^-(0)$ & $\frac{V^-(0)}{G}$ & $\frac{1}{G}$ & $\max[G]$ & Maximize gain by increasing pump power, length, or through the use of resonances. \\ 
 
 \hline
 Internal Loss & Loss within the squeezer injects vacuum noise & Loss rate $\alpha$, gain rate $g$ & $\frac{\alpha}{\alpha+g}$ & $\frac{\alpha}{g}$ & $L \approx \frac{\ln(\frac{g}{\alpha}) }{2g}$  & Increase pump power, employ pump resonance, select platform with strong nonlinearity, and minimize propagation losses in order to maximize the nonlinearity to loss ratio. \\ 

  \hline
Pump Saturation & Amplified light acquires pump noise & Amplified seed $P_1^+$, pump power $P_3$, pump variance $V^-_3$ & $\frac{ P_1^+}{2 P_3}V^-_3$ & $\sqrt{\frac{P^+_1(0)}{P_0}}$ & $G = \sqrt{\frac{2P_0}{P^+_1(0)}}$ & Minimize fundamental power sent through squeezer along amplified quadrature by filtering power or adjusting its phase. \\

  \hline
External Loss & Loss after the squeezer injects vacuum noise & transmission $\eta_e$ & $1-\eta_e$ & $1-\eta_e$ & $\max[\eta_e]$& Minimize loss in squeezed light path. \\

  \hline
Phase Noise & Proper phase needed to measure squeezed quadrature & rms phase deviation $\delta\theta$ & $G\sin^2\delta\theta$ & $\sin 2\delta\theta$ & $G = \cot \delta\theta$ &  Lock squeezed to reference phase to mitigate phase drift, and optimize feedback loop gain, bandwidth. \\

  \hline
Intensity Noise & Intensity noise in the reference beam, if not properly suppressed or if the squeezed beam has substantial power, adds noise & Reference intensity noise $V^X_\text{ref}$, imbalance $1-\mathcal{V}_\text{imbalance}^2$, power $P^\pm_{1}$ in initial anti-squeezed and squeezed quadratures, reference powers $P_\text{LO}, P_\text{3,det}$ for BHD and PHD & 
$ \substack{
V^X_{\mathrm{LO}} (1-\mathcal{V}_\text{imbalance}^2) 
 +\frac{ P_1^- V^X_{\mathrm{LO}} + P_1^+ V^Y_{\mathrm{LO}} }{P_{\mathrm{LO}}}
,  \ \text{BHD}  \\
     \frac{2\bigg(\frac{\partial G_{\mathrm{det}}}{\partial P}|_{P_\text{3, det}}\bigg)^2 P_1^- P_\text{3, det}  V_\text{3,det}^X }{G_\text{det}}  
 ,  \  \text{PHD} }
$
 & $\frac{P_1}{P_\text{ref}}$ & $\min[\frac{P_1}{P_\text{ref}} V^X_\text{ref}]$ &  Use balanced detection, filter reference beam or reduce signal power. Increase reference power if its contribution is shot-noise limited. \\

\hline
Detection Loss & Loss and mode mismatch inject vacuum noise & Detection efficiency $\eta_{\mathrm{det}}$ (includes mode matching, detector quantum efficiency)&
$\substack{ 1-\eta_{\mathrm{det}, \ \ \text{BHD}}\\
\frac{\alpha}{g_\text{det} - \alpha} + \frac{1-\eta_{\mathrm{det}}}{G_\text{det}},  \ \ \text{PHD} }$ 
& $\substack{ 1-\eta_{\mathrm{det}} \\ \frac{\alpha}{g_\text{det}}  + \frac{1-\eta_{\mathrm{det}}}{G_\text{det}} }$ 
& $\substack{ \max[\eta_{\mathrm{det}}] \\ , \max[\frac{\alpha}{g_\text{det}}, G_\text{det}] }$ & Optimize mode matching and coupling to detector, reduce reflections, increase absorption in detector, use high-gain detection amplifier.\\
 
\hline
 Electronic Noise & Detected power should be sufficient so that shot noise clears the electronic noise floor & Detector noise-equivalent power $\text{NEP}$, quantum efficiency $\eta_{\mathrm{QE}}$, detected power $P_\text{det}$ & 
 $\substack{ \frac{\text{NEP}^2 \eta_\text{QE} }{2 h \nu} \frac{1}{P_\text{det}}, \ \ \text{BHD}\\
\frac{\text{NEP}^2 \eta_\text{QE} }{2 h \nu} \frac{1}{P_\text{det} G_\text{det}},  \ \ \text{PHD} }$  
 &  $\substack{ \frac{\text{NEP}^2 \eta_\text{QE} }{2 h \nu P_\text{det}} \\
\frac{\text{NEP}^2 \eta_\text{QE} }{2 h \nu P_\text{det} G_\text{det}} }$ & $\substack{ \min[\frac{\text{NEP}^2  }{ P_\text{det}}]  \\ \max[G_\text{det}] }$ & Increase detected power, use low-noise electronics, use high-gain detection amplifier. \\ [1ex] 
 \hline
\end{tabular}
\caption{\label{summary_table} \textbf{Summary of Noise Limits on Squeezing.}}
% \end{sidewaystable}
\end{table}
% \end{center}
\end{landscape}

%%%%%%%%%%%%%%%%%%%%%%%%%%%%%%%%%%%%%%%%%%%%%%%%%%%%%%%%%%%%%%%%%%%%%%%%%
% \section*{Prospects}
%%%%%%%%%%%%%%%%%%%%%%%%%%%%%%%%%%%%%%%%%%%%%%%%%%%%%%%%%%%%%%%%%%%%%%%%%

Table \ref{ref_table} compares squeezing performance and noise bounds of different integrated continuous wave squeezing systems to date. Squeezing bounds were calculated from reported numbers. Note that the limiting noise term for most of the squeezers cited is the combination of external and detection loss. For comparison, the record bulk squeezer reference has been included.

\begin{table}[h!]
\footnotesize
\hspace*{-0.05\linewidth}\begin{tabular}{||m{0.02\textwidth} m{0.07\textwidth} m{0.07\textwidth} m{0.06\textwidth} m{0.05\textwidth} m{0.03\textwidth} m{0.06\textwidth} m{0.04\textwidth} m{0.06\textwidth} m{0.04\textwidth} m{0.06\textwidth} m{0.06\textwidth} m{0.07\textwidth} 
||} 
 \hline
 Ref. & Platform & Resonance & Detection & Pump Power (mW) & Gain (dB) & External + Detection Loss Bound (dB) & Internal Loss Bound (dB) & Pump Saturation Bound (dB) & Phase Noise Bound (dB) & Intensity Noise (dB) & Electronic Noise (dB) & Measured Squeezing (dB) \\ [0.5ex] 
 \hline

\hline
\cite{hirota2025} & LN & None & BHD & 640 & 20 & -12 & -17 & - & -20 & - & -28 & -10 \\ 
 
\hline
\cite{Furusawa2023} & LN & None & BHD & 660 & 20 & -9.2 & -21 & - & -16 & - & -25 & -8.3 \\ 

\hline
\cite{Kashiwazaki2021} & LN & None & PHD & 490 & 17  & -6.8 & -17 & - & - & - & - & -6.3 \\ 

\hline
\cite{Inoue2023} & LN & None & pBHD & 438 & 15  & -5.9 & -17 & - & -26 & - & -27 & -5.2 \\ 

\hline
\cite{Nehra2022} & (Pulsed) TFLN & None & PHD & - & 10 & -5.2 & - & - & - & - & - & -4.9
\\ 

\hline
\cite{Shi2025} & TFLN & None & BHD & 38 & 5 & -3.9 & - & - & - & - & - & -1.4
\\ 

\hline
\cite{Park2024} & TFLN & Signal & BHD & 4 & 4.4 & -2.8 & -4.6 & - & - & - & -5.2 & -0.55 \\

\hline
\cite{Andersen2024} & TFLN & Signal, Pump & BHD & 6.9 & 6.8 & -1.2 & -3.5 & - & - & - & - & -0.46
\\ 
 \hline
\cite{Stokowski2023} & TFLN & None & BHD & 3.4 & 1.5 & -1 & -12 & -9 & - & - & -10 & -0.12 \\

\hline
\cite{Ulanov2025} & SiN & Signal, Pump & BHD & 33 & 9.9 & -4.3 & -10 & - & - & - & -20 & -1.71 \\

\hline
\cite{Vernon2021} & SiN & Signal, Pump & BHD & 70 & 8 & -2.1  & -10 & - & - & - & -14.7 & -1.65 \\

\hline 
\cite{Gaeta2020} & SiN & Signal, Pump & BHD & 102 & 10.8 & -3.2 & -4.8 & -14 & - & - & -4 & -0.81 \\

\hline
\cite{Politi2020} & SiN & Signal, Pump & Pol. HD & 52 & 5 & -3.3 & -6.4 & - & - & - & -6 & -0.45 
\\ 

\hline
\cite{Purdy2013} & SiN & Optics, Mechanics & BHD & 0.0002 & - & -1.6 & -2.2 & - & - & - & - & -1.7
\\ 

\hline
\cite{Safavi-Naeini2013} & Si & Optics, Mechanics & BHD & 0.004 & - & -1.3 & -3.5 & - & - & - & -15 & -0.2
\\ 

\hline
\cite{Schnabel2016} & (Bulk) PPKTP & Signal, Pump & BHD & 16 & 25 & -16 & -20 &  & -25 & - & -28 & -15.3 
\\ [1ex] 
 \hline

\end{tabular}
 \caption{\label{ref_table} \textbf{Literature survey of (mostly) continuous wave integrated squeezers, grouped by platform.} Noise contributions were extracted from reported numbers, where available. BHD - balanced homodyne detection. PHD - parametric homodyne detection. pBHD - preamplified balanced homodyne detection. Pol. HD - polarization homodyne detection.}

\end{table}

Integrated squeezers show promise for low-power squeezed light systems - most of the pump powers used in the integrated squeezers above range from few to hundreds of mW. From table \ref{ref_table}, we see that multiple platforms show potential for integrated squeezers, such as $\chi^{(2)}$ mechanically polished lithium niobate, $\chi^{(2)}$ thin-film lithium niobate and $\chi^{(3)}$ low-loss silicon nitride. Each platform offers distinct advantages and disadvantages - mechanically polished lithium niobate waveguides provide efficient outcoupling but it can be difficult to fabricate waveguide bends\cite{Kashiwazaki2021}, thin film lithium niobate waveguides offer large nonlinear rates but less efficient outcoupling\cite{Stokowski2023}, and silicon nitride waveguides offer ultra-low-losses and more established fabrication but also lower nonlinearities\cite{Gaeta2020}. Although balanced homodyne detection has been the dominant detection method for squeezed light, we expect further developments in parametric homodyne detection and optical preamplification to capitalize upon the noise-tolerant benefits of phase-sensitive amplification.
Most of the squeezers were limited by external and detection loss, with the dominant roadblock arising from coupling losses to off-chip photodetectors. Efficient chip-fiber coupling designs is an active research area with greater than $96\%$ coupling achieved in silicon nitride \cite{Zhu2016} and greater than $88\%$ achieved in thin film lithium niobate\cite{Hu:21}. Note that fully on-chip squeezed light systems could utilize parametric homodyne detection or on-chip photodetection to mitigate or circumvent the effect of outcoupling losses \cite{Nehra2022, PIELS20163, Ahn:23}.  

Demonstrations in lithium niobate waveguides have achieved waveguide loss rates less than 3 dB/m \cite{Loncar2017, Loncar2022, Khalat2025} and gain rates exceeding 40 dB/cm with pulses \cite{Ledezma2022} and 20 dB/cm with continuous wave systems \cite{Dean2026}. The internal loss bound on squeezing can therefore be well below -20 dB. Future squeezer devices should optimize for nonlinearity-to-loss ratio, rather than solely low loss (achieved with wide waveguides\cite{Loncar2022, Khalat2025}) or high nonlinearity (achieved with small waveguides). Pump saturation effects have not been reported in most experiments. Even if they do appear at larger gains, they can be avoided through effective filtering and mitigation of reflections. Similarly, phase noise has not limited integrated experiments so far and we expect the small length scales and inherent stability of chip-scale photonics will aid in mitigating phase noise. Intensity noise is often avoided by choosing measurement frequency bands where it is negligible, and isn't commonly reported. However, as more squeezing is pursued, we expect that clearance over excess reference noise will become an important metric. The electronic noise can be made very small, with some experiments measuring as much as 28 dB noise clearance over the shot noise \cite{Schnabel2016, hirota2025}.

All put together, the noise challenges listed in this paper are surmountable and we expect substantial squeezing in excess of 15 dB will be achieved in integrated squeezers in the coming years, opening the door for many new applications with squeezed light. High levels of squeezing will improve certain precision sensors - detecting weaker signals faster\cite{LIGO2020, Bowen2021} - and also quantum information processing - enabling quantum computation \cite{Madsen2022}. Even before squeezed light can be used in real-life sensing and computing systems, the coming squeezer measurements will provide detailed measurements of noise, such as the quantum efficiency of efficient photodiodes \cite{Schnabel2016}. To quote Levensen and Shelby’s insightful 1986 outlook on bulk squeezers, “The next generation of experiments will be designed to avoid these noise sources. The future will show whether the result will be many dB of noise suppression or the discovery of new sources of optical noise” \cite{Levenson1986}.

%%%%%%%%%%%%%%%%%%%%%%%%%%%%%%%%%%%%%%%%%%%%%%%%%%%%%%%%%%%%%%%%%%%%%%%%%
% \section*{Conclusion}
%%%%%%%%%%%%%%%%%%%%%%%%%%%%%%%%%%%%%%%%%%%%%%%%%%%%%%%%%%%%%%%%%%%%%%%%%
In this paper, we derived a general analytic model for optical squeezers. We summarized well-known noise limits, derived simple equations and intuition for some less well-defined limits, and combined all the limits in a simple and useful way. We then discussed the implications on squeezer design and simple measurements to enable design iteration. Lastly, we commented on the current state-of-the-art and future prospects for integrated squeezers. 

%%%%%%%%%%%%%%%%%%%%%%%%%%%%%%%%%%%%%%%%%%%%%%%%%%%%%%%%%%%%%%%%%%%%%%%%%%%%%%%%%%%%%%%%%%%%%%%%%%%%%%%%%%%%%%%%%%%%%%%%%%%%%%%%%%%%%%%%%%%%%%%%%%%%%%%%%
\section*{Methods}

%%%%%%%%%%%%%%%%%%%%%%%%%%%%%%%%%
\subsection*{Theoretical Framework} \label{sec:methods_gaussian}

The general quantized electric field $\hat E(t)$ of monochromatic light can be expressed as 

\begin{equation}
\hat E(t) =  \hat a(t)e^{-i\omega_0 t},
\end{equation}
where $\hat a (t)$ is the slowly varying field annihilation operator and the rapid oscillations are at frequency $\omega_0$. For the typical case where quantum fluctuations are relatively small compared to the mean field, we can linearize the field operator into a classical mean field $A(t)$ and a quantum fluctuation operator $\delta \hat{a}$:
\begin{equation}
\hat a(t) = A(t)+\delta\hat{a}. 
\end{equation}
This approach falls within the framework of Gaussian quantum optics, as the linearization of field operators reduces the order of the three- or four-wave interaction and leads to quadratic Hamiltonians of the form $\hat{H} \propto \hat{a}^{\dagger}\hat{a}$, $\hat{a}^{2} + \hat{a}^{\dagger 2}$, or $\hat{a}\hat{b}^{\dagger} + \hat{a}^{\dagger}\hat{b}$. A fundamental property of these quadratic Hamiltonians is that they preserve the Gaussian character of quantum states. If the initial state is Gaussian in phase space, it remains Gaussian after evolution under such Hamiltonians\cite{weedbrookRMP,Jankowski2024}. 

We define the quadrature fluctuation operators by 
\begin{align}
 \hat{X} &= 2\operatorname{Re}[\hat{a}(t)] = \hat{a} + \hat{a}^{\dagger}\\
\hat{Y} &= 2\operatorname{Im}[\hat{a}(t)] = -i(\hat{a} - \hat{a}^{\dagger})
\end{align}
with commutation relation 
$[\hat X, \hat Y] = 2i$ , which implies the Heisenberg uncertainty product 
\begin{equation} 
V^X V^Y \ge 1,
\end{equation}
where $V^X$ and $V^Y$ are the variances in the $\hat X$ and $\hat Y$ quadratures defined as
\begin{align}
V^X &\equiv \langle(\hat{X} - \langle\hat{X}\rangle)^2\rangle = \langle\hat{X}^2\rangle - \langle\hat{X}\rangle^2\\
V^Y &\equiv \langle(\hat{Y} - \langle\hat{Y}\rangle)^2\rangle = \langle\hat{Y}^2\rangle - \langle\hat{Y}\rangle^2.
\end{align}

%%%%%%%%%%%%%%%%%%%%%%%%%%%%%%%%%%
\subsection*{Noise Mixing Approximation} \label{sec:App:lowlossapprox}

We stated in the introductory section that, in the limit of small noise mixing terms, the usual beamsplitter expression that describes mixing noise sources becomes a simple additive expression.
Since this additive expression is the backbone of the paper, we take some time here to more concretely define it and its limitations.

Noise processes affect the variance $V$ according to
\begin{equation}
V_{\text{out}} = \eta V_{\text{in}} + (1-\eta)V_{\rm noise},
\end{equation}
where $\eta$ is the transmission efficiency ($0 \leq \eta \leq 1$) and $V_{\rm noise}$ is the variance of the added noise (typically vacuum fluctuations), which is assumed to be uncorrelated with the signal. For a sequence of $j$ such processes, the final variance $V_j$ is obtained by applying this relation sequentially. 
To see how this simplifies, consider the case of just two loss sources:
\begin{align}
    V_2 &= \eta_2 V_1 + (1-\eta_2)V_{\text{noise},2} \nonumber \\
    &= \eta_2 [\eta_1 V_0 + (1-\eta_1)V_{\text{noise},1}] + (1-\eta_2)V_{\text{noise},2} \nonumber \\
    &\approx V_0 + (1-\eta_1)V_{\text{noise},1} + (1-\eta_2)V_{\text{noise},2} \\
    &\approx V_0 + \Delta V_\text{noise,1} + \Delta V_\text{noise,2}
\end{align}
where in the last line we assumed low loss ($1-\eta \ll 1$) such that the signal attenuation is negligible ($\eta_1\eta_2 \approx 1$) and the noise added in the first step is not significantly attenuated by the second step ($\eta_2(1-\eta_1) \approx 1-\eta_1$).
Generalizing this to $j$ steps yields the simple sum:
\begin{align}
V_j &= \eta_j V_{j-1} + (1-\eta_j)V_{{\rm noise},j} \label{eq_noise_recursive}\\
&\approx V_0 + \sum_{i=1}^j (1-\eta_i) V_{{\rm noise},i},  \ \ \ \ {1-\eta_i \ll1, \forall i} \\
    &\approx V_0 + \sum_{i=1}^j \Delta V_\text{noise,i} \label{eq_noise_sum} 
\end{align}
where the first line defines the exact recursive evolution and the second line gives the additive approximation.

\begin{figure}
    \centering
    \includegraphics[width=\linewidth]{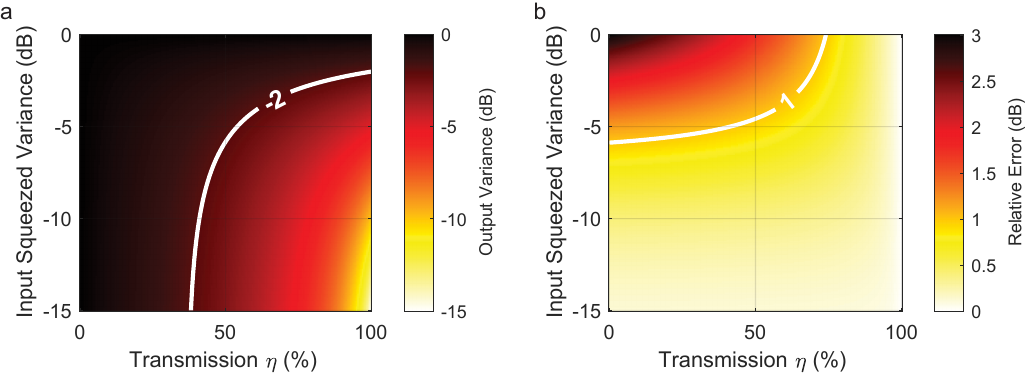}
    \caption{ \label{fig:low_loss_approx_fig} \textbf{Noise summation approximation validity.} (a) Output variance as a function of transmission and input variance. Contour shows the $-2$ dB output variance line, below which large squeezing is achieved. (b) Relative error of the small noise-mixing approximation as a function of transmission and input variance. Contour shows the $1$ dB relative error line, below which the approximation is valid (within 1 dB of the true value).}
\end{figure}

In this section, we explicitly derive the range of loss and noise mixing terms where the approximation is valid. Consider the approximation of a single noise process
\begin{align}
V_{\text{out}} &= \eta V_{\text{in}} + (1-\eta)V_{\rm noise}, \label{eq:low_loss_exact_eq}\\
&\approx V_{\text{in}} + (1-\eta)V_{\rm noise}.
\end{align}
The relative error of the approximation is 
\begin{align}
\frac{ V_\text{approx} - V_\text{out}}{V_\text{out}} = \frac{(1-\eta)V_\text{in}}{\eta V_{\text{in}} + (1-\eta)V_{\rm noise}}.\\
\end{align}
Figure \ref{fig:low_loss_approx_fig}a plots the output variance of equation \ref{eq:low_loss_exact_eq} as a function of both transmission and input squeezing.
Figure \ref{fig:low_loss_approx_fig}b plots the relative error of the approximation for squeezing ($V_\text{in} \leq 1$) mixed with noise ($V_\text{noise} \geq 1$) and shows that the relative error is smallest for small noise mixing terms $1-\eta \ll 1$ and for small input variance $V_\text{in}\ll 1$. More explicitly, the relative error is smaller than 1 dB ($\pm20\%$) for any squeezing measurement that detects more than 2 dB of squeezing. So despite being referred to in the main text as a "low noise-mixing" approximation, it is in fact very robust and applies not just to highly squeezed systems but also moderately squeezed ones. 
Furthermore, the relative error is always positive which indicates the approximation provides is conservative  - it consistently underestimates the squeezing.

%%%%%%%%%%%%%%%%%%%%%%%%

%%%%%%%%%%%%%%%%%%%%%%%%%%%%%%%%%%%%%%%%
\subsection*{Internal Loss} \label{app:internalloss}

A key result from our work is the universal limit of internal loss on squeezers and detection amplifiers.
In this section, we introduce our framework for internal loss and show that in the different squeezers and detection amplifiers it reduces effectively to the inverse of the nonlinearity to loss ratio.

\subsubsection*{Nonresonant Squeezer Internal Loss} 
Squeezing and loss happen simultaneously within the squeezer itself. We can expand this in a split-step manner to be infinitesimal segments $\Delta L$ of pure squeezing followed by infinitesimal segments of pure loss as illustrated in Figure~\ref{fig:loss}c.

\begin{align*}
V(L+\Delta L) &= \eta(\Delta L)\, V(L)\, S(\Delta L) + \left(1-\eta(\Delta L)\right) \\
&= e^{-2\alpha \Delta L} e^{-2g\Delta L} V(L) + \left(1-e^{-2\alpha \Delta L}\right) \\
&= \left(1-2\alpha \Delta L - 2g\Delta L\right)V(L) + 2\alpha \Delta L, \\
\frac{V(L+\Delta L) - V(L)}{\Delta L}
&\approx \frac{\partial V}{\partial L}
= -2(\alpha + g)V + 2\alpha .
\end{align*}
We obtained a simple differential equation for how squeezing evolves. We can solve this using the substitution
\[
\tilde V = V e^{2(\alpha+g)L}, \qquad
\frac{\partial \tilde V}{\partial L} =
\left[\frac{\partial V}{\partial L} + 2(\alpha+g)V\right] e^{2(\alpha+g)L}
\]
to arrive at
\begin{align*}
\frac{\partial V}{\partial L} &= -2(\alpha+g)V + 2\alpha, \\
\tilde V &= V e^{2(\alpha+g)L}, \\
\frac{\partial \tilde V}{\partial L} &=
\left[\frac{\partial V}{\partial L} + 2(\alpha+g)V\right] e^{2(\alpha+g)L}
= 2\alpha e^{2(\alpha+g)L}.
\end{align*}
Solving the differential equation, we find
\[
\tilde V(L) - V(0)
= \int_0^L 2\alpha\, e^{2(\alpha+g)l} \, dl
= \frac{2\alpha}{2(\alpha+g)}\left(e^{2(\alpha+g)L} - 1\right).
\]

\[
V(L)
= e^{-2(\alpha+g)L}
\left[
\int_0^L 2\alpha\, e^{2(\alpha+g)l} dl + V(0)
\right],
\]

\[
V(L)
= \frac{\alpha}{\alpha+g}\left(1-e^{-2(\alpha+g)L}\right)
+ e^{-2(\alpha+g)L} V(0).
\]

\[
V(L) \approx
\begin{cases}
e^{-2gL} V(0), & L \le L_0 \quad \text{loss-tolerant}, \\
\dfrac{\alpha}{\alpha+g}, & L > L_0 \quad \text{loss-limited}.
\end{cases}
\]

\[
V(L) \approx e^{-2gL}V(0) + \frac{\alpha}{\alpha+g}, \qquad g \gg \alpha.
\]
We see that the maximum squeezing attainable is nothing but the ratio of loss and nonlinearity rates,
\[
\frac{\alpha}{\alpha + g}.
\]
In the limits of pure squeezing and pure loss we recover the expected values:
\[
\begin{aligned}
V(L)
&= e^{-2(\alpha+g)L}
\left[
\int_0^L 2\alpha\, e^{2(\alpha+g)l}dl + V(0)
\right] \\
&=
\begin{cases}
e^{-2\alpha L}(e^{2\alpha L}-1) + e^{-2\alpha L}V(0), & g\to 0, \\
e^{-2gL} V(0), & \alpha\to 0.
\end{cases}
\end{aligned}
\]
The boundary between regimes occurs when the loss-limited squeezing intersects the loss-tolerant squeezing:
\[
\frac{1}{1 + \frac{e^{2\alpha L_0} - 1}{V_0}}
\equiv 
\frac{1}{1 + \frac{\alpha}{g-\alpha}\frac{1}{V(0)} }.
\]
\[
e^{2\alpha L_0} = g,
\qquad
L_0 \equiv \frac{\ln(g)}{2\alpha}.
\]
Equivalently, the transition between loss-limited and loss-tolerant squeezing occurs at the boundary of the two asymptotes:
\[
e^{-2gL_0}
\equiv
\frac{\alpha}{\alpha+g},
\qquad
L_0 = \frac{\ln\!\left(\frac{\alpha+g}{\alpha}\right)}{2g}.
\]
Using the definition of internal loss given in Eq. \eqref{eq:internal_loss_def}, the internal loss of a nonresonant squeezer is 
\begin{align}
\Delta V_\text{internal loss, nonres}  &= 1- \frac{\frac{\alpha}{\alpha+g} [1-e^{-2(\alpha+g)L} ] + e^{-2(\alpha+g)L} - 1}{e^{-2gL} - 1}\\
\Delta V_\text{internal loss, nonres}  &\approx \begin{cases}
 1-e^{-\alpha L} & L\leq L_{\eta_i}\\
 \frac{\alpha}{\alpha+g} & L>L_{\eta_i}  
\end{cases}
,  \ \ \ \ \text{ where } L_{\eta_i} \equiv \frac{1}{\alpha+g}
\end{align}

The fact that large squeezing can be obtained despite the presence of simultaneous loss was noted in Ng et al.\cite{Ng2023}, where a squeezed variance of $V=0.6\%$ (-22 dB) was simulated despite a total loss of $1-\eta= 4\%$ (-14 dB) distributed throughout the squeezer.
Note that the nonlinearity-to-loss ratio discussed here is related to, but distinct from, the single photon nonlinearity to loss ratio “$g_{sp}/\kappa_{a,i}$” mentioned in the context of single photon nonlinearity \cite{Tang2020, Fang2022, Yanagimoto2022, Jankowski2024}. Specifically, the nonlinearity $g$ we refer to includes a pump photon number much larger than one. However, many of the conclusions of literature on how to maximize the single-photon nonlinearity to loss ratio also apply to maximizing the total nonlinearity to loss ratio.

\subsubsection*{Signal-resonant Squeezer Internal Loss}
In this section we re-express the squeezing out of a signal-resonant squeezer in terms of the internal loss defined in the main text.
The threshold condition (unity round-trip gain) in terms of average field spatial gain rate $g_\text{th}$, resonator field outcoupling $c$, average field spatial loss rate $\alpha$, and the length $L$ is
\begin{align}
    1 &= e^{2g_\text{th}L} (1-c^2)e^{-2\alpha L} \\
    0 &= 2g_\text{th}L  -2\alpha L + \ln( 1-c^2)\\
    g_\text{th} &= \alpha - \frac{\ln(1-c^2)}{2L}
\end{align}
and the extraction efficiency (outcoupling rate to total loss rate) is 
\begin{align}
    \rho &= \frac{c^2}{1-e^{-2\alpha L}(1-c^2)} \\  
\end{align}
Assuming a low loss cavity $ 2\alpha L \ll 1, c^2 \ll 1$, we find  
\begin{align}
    g_\text{th} &\approx \alpha + \frac{c^2}{2L}, \\
    \rho &\approx \frac{ (g_\text{th}-\alpha) 2L } {2\alpha L + (g_\text{th}-\alpha) 2L} = \frac{g_\text{th}-\alpha}{g_\text{th}}. \label{eq:g_th_deriv}
\end{align}
The squeezing in a below-threshold signal-resonant squeezer is bounded by the extraction efficiency
\begin{align} 
    V^- &= 1-\rho \frac{4 \frac{g}{g_\text{th}}}{(1+\frac{g}{g_\text{th}})^2} \\
    &> 1-\rho , & \frac{g}{g_\text{th}} < 1. \label{eq:signal_res_Vs_deriv}
\end{align}
Combining Equations \ref{eq:signal_res_Vs_deriv} and \ref{eq:g_th_deriv} give the final bound
\begin{align}
    V^- > \frac{\alpha}{g_\text{th}}.
\end{align}

%%%%%%%%%%%%%%%%%%%%%%%%%%%%%%%%%%%%%%%%%%%%%%%%%%%%%%%%%%%%%%%%%%%%
\subsubsection*{Nonresonant Parametric Homodyne Detection Internal Loss}

In parametric homodyne detection, both the signal and noise get exponentially amplified. In the presence of loss, a small amount of vacuum noise is added near the beginning of the amplifier where the state is still sensitive to loss. Our amplifying differential equations for variance and mean power are 
\begin{align}
\frac{\partial V}{\partial z} &= -2(\alpha - g)V + 2\alpha, \\
\frac{\partial P}{\partial z} &= -2(\alpha - g)P,
\end{align}
which have solutions
\begin{align}
V(L) &= \frac{\alpha}{g-\alpha}\!\left[e^{2(g-\alpha)L}-1\right] + e^{2(g-\alpha)L} V(0), \\
P(L) &= e^{2(g-\alpha)L} 
\end{align}
Let us express the result in terms of the effective internal loss that would degrade squeezing the same way. Recall internal loss is defined as
\begin{align}
V(L) &= \eta_i\, V_{\mathrm{ideal}} + (1-\eta_i), \\
\eta_i &= \frac{V(L)-1}{V_{\mathrm{ideal}} - 1}.
\end{align}
We already know that the variance evolves as
\[
V(L) = \frac{\alpha}{g-\alpha}\!\left[e^{2(g-\alpha)L}-1\right] + e^{2(g-\alpha)L}V(0),
\]
and in an ideal amplifier,
\[
V_{\mathrm{ideal}} = e^{2gL}V(0).
\]
Therefore the internal loss becomes

\begin{align}
\eta_i &=
\frac{
\frac{\alpha}{g-\alpha}\!\left[e^{2(g-\alpha)L}-1\right] + e^{2(g-\alpha)L}V(0) - 1
}{
e^{2gL}V(0) - 1
}, \\
1 - \eta_i &=
\frac{
e^{2gL}V(0) - 1 -
\left[
\frac{\alpha}{g-\alpha}\!\left[e^{2(g-\alpha)L}-1\right]
+ e^{2(g-\alpha)L}V(0) - 1
\right]
}{
e^{2gL}V(0) - 1
}.
\end{align}
This expression looks complicated, but it is not fundamentally different from what we found earlier. We can interpret anti-squeezing as time-reversed squeezing with negative loss rate. From the expression for $V(L)$ we see that the first term is the noise term, and it quickly approaches the constant
\[
\frac{V(L)}{G_{\mathrm{PHD}}} \longrightarrow \frac{\alpha}{g-\alpha},
\]
while for small lengths, ordinary loss contributes the usual amount of noise. We therefore expect the internal loss to evolve as
\[
\eta_i \approx
\begin{cases}
e^{-2\alpha L}, & L \le L_\eta, \\
\displaystyle \frac{g - 2\alpha}{g - \alpha}, & L > L_\eta,
\end{cases}
\]
and we can therefore bound the internal loss by
\[
\boxed{
1 - \eta_i \;\lessapprox\; \frac{\alpha}{g-\alpha}.
}
\]

%%%%%%%%%%%%%%%%%%%%%%%%%%%%%%%%%%
\subsection*{Applying and Visualizing the Model}
\label{sec:App:preampBHD}

In this section, we apply our model to estimate the noise limitations on an example system. Our aim is to illustrate how to apply our model and also visualize the different noise contributions for a realistic system. We link to a Google Colab Notebook that can help plot and visualize the dependence of noise on various parameters.  

Our example system consists of a squeezer composed of a nonresonant $\chi^{(2)}$ optical parametric amplifier and detection that consists of balanced homodyne detection with an optical preamplifier (another nonresonant $\chi^{(2)}$ optical parametric amplifier) to provide loss and noise tolerance.
Applying our master equation \eqref{eq:full_noise_model} to this specific squeezed light system, we arrive at 
\begin{align} \label{eq:ex_full_model}
    V &\approx 
\frac{1}{G} + \frac{\alpha}{g} 
 + \frac{P_1^+}{2P_3}V_3^-  + (1-\eta_{e}) +  \frac{G\sin^2\delta\theta}{G_\text{det}} \\ \nonumber
    &+  \frac{1}{2}(\ln G_{\mathrm{det}})^2\frac{ P_1^-}{P_3}V_\text{3,det}^X 
 +  \frac{\alpha}{g_\text{det}}   \\ \nonumber
    &+  \frac{V^X_{\mathrm{LO}} (1-\mathcal{V}_\text{imbalance}^2) + 
 \frac{ P_1^- V^X_{\mathrm{LO}} + P_1^+ V^Y_{\mathrm{LO}} }{P_{\mathrm{LO}}} + (1-\eta_{\mathrm{det}})  + \frac{\text{NEP}^2 \eta_\text{QE} }{2 h \nu P_{\mathrm{LO}}}   }{G_\text{det}}, 
\end{align}
where the first line contains the squeezed light generation noise terms, the second line contains noise and loss terms from the preamplification process, and the third line contains the balanced homodyne noise terms, reduced by the preamplifier gain. Note that the phase noise contribution is also reduced by the preamplifier gain, since the measured antisqueezing is reduced by the preamplifier. Power and phase fluctuations on the preamplifier pump both can contribute to intensity noise on the amplified signal, though the effect of phase fluctuations disappears to first order if the phase is properly biased, so for simplicity here we assume only the power fluctuations are significant (term 6 of equation \eqref{eq:ex_full_model}).

\begin{figure}
    \centering
    \includegraphics[width=\linewidth]{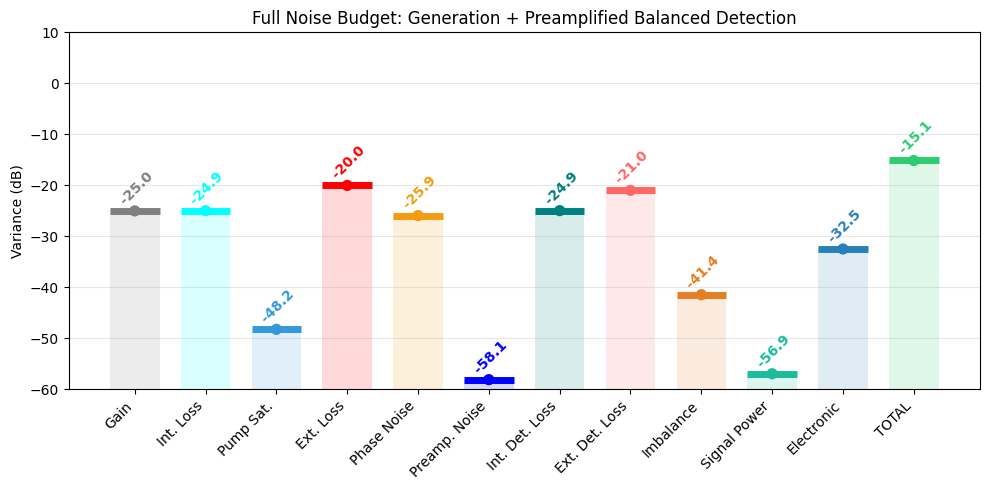}
    \caption{ \label{fig:fullmodel_fig} \textbf{Example noise contribution visualization with model.} To see the exact parameters used in the simulation and to test various parameter values, see the link to the code.}
\end{figure}

We now input reasonable parameter values based on currently demonstrated devices. Dean et al. demonstrated an optical parametric amplifier with high gain $G \approx 20$ dB for an input power of $P_\text{in} \approx 300$ mW, high nonlinearity-to-loss ratio $\frac{g}{\alpha} \approx 20$ dB, and couplers capable of very high extraction efficiency $\eta_e \approx 0.99$ and large pump filtering $\frac{P_1}{P_\text{in}} < -50$ dB \cite{Dean2026}.
Hirota et al. measured phase noise as low as $\delta\theta \approx 9$ mrad, electronic noise clearance over 28 dB at local oscillator power $P_\text{LO}\approx 15$ mW, and detection efficiency as high as $\eta_\text{det}\approx 0.92$\cite{hirota2025}. 
Using these numbers in our model and further assuming $G_\text{det} = 10$ dB preamplifier gain and reasonable numbers for classical laser relative intensity noise $\text{RIN}_\text{cl} \approx -160$ dBc/Hz for both local oscillator and pump, and that the leakage phase is set so that it is deamplified in the squeezer $P_1^+ \ll P_1^-$, that the imbalance is typical $\mathcal{V}^2_\text{imbalance} \approx -50$ dB, we find that as much as 15 dB of squeezing is feasible.
Figure \ref{fig:fullmodel_fig} displays the various noise contributions in this example calculation. For more information and to test different parameter values in the code, see the GUI at \href{https://colab.research.google.com/drive/13sn-cy8ZQcc7aY9DQnGOqcuVF4GuwHrD?usp=sharing&authuser=1#scrollTo=pyQ-H2_EhAC6}{Google Colab Notebook Link}.

%%%%%%%%%%%%%%%%%%%%%%%%%%%%%%%%%%%%%%%%%%%%%%%%%
\section*{Data Availability}
The datasets generated and/or analyzed during the current study are not publicly available due to their nature as numerical evaluations of the analytical models presented in the text, but are available from the corresponding author on reasonable request.

\section*{Code Availability}
We provide an interactive plotting script, for plotting and visualizing the effects of different parameters on noise contributions and total noise budget, through Google Colab linked in the Methods section "Applying and Visualizing the Model". The calculation code that supports the findings of this study are available from the corresponding authors upon request.

\section*{Acknowledgements}         
\vspace{-2mm}
This work was supported in part by the Defense Advanced Research Projects Agency (DARPA) INSPIRED program (HR00112420356).
We also thank NTT Research for their financial and technical support. D.D. acknowledges support from the NSF GRFP (No. DGE-1656518). H.S. acknowledges support from the Urbanek Family Fellowship. We are grateful for insightful discussions with Dr. Justin Cohen at DARPA, Dr. Edwin Ng at NTT Research and Mr. Kevin Multani.  
\vspace{-2mm}

\section*{Author Contribution}
All authors contributed to the insights into the noise limits in integrated squeezers.
In addition, D.J.D. collated the noise contributions into one model and derived the contributions due to internal loss and pump saturation, as well as bounds for the other noise terms.
T.P. and S.R. contributed to sections on signal-resonant systems, detection-related noise sources, and phase noise.
L.S.M., A.T., J.J.S., and W.P.B. performed numerical simulations and calculations regarding phase noise, thermorefractive noise, pump saturation, external and internal loss, and imbalanced homodyne detection. 
G.H.A and Z.W. contributed to the intensity noise section.
L.Q. contributed to the estimates of loss section.
H.S. contributed to understanding the pump saturation and losses.
D.S. contributed to the internal loss derivation for nonresonant squeezers as well as the estimation of detector quantum efficiency.
J.C., D.S., W.P.B., M.M.F., and A.H.S-N. supervised the project and provided subject-matter expertise.
D.J.D. and A.H.S-N. wrote the manuscript with input from all authors.

\section*{Competing Interests}
The authors declare no competing non-financial interests but the following competing financial interests. D.J.D., T.P., H.S., and A.H.S.-N. are inventors on filed patent application PCT/US2025/44973 filed by The Board of Trustees of the Leland Stanford Junior University that covers methods for achieving quantum advantage in power-constrained photonic sensors, which is related to the generation and detection of squeezed light discussed throughout the manuscript. D.J.D., G.H.A, Z.W., and A.H.S-N. are inventors on filed patent disclosure application 63/916962 filed by The Board of Trustees of the Leland Stanford Junior University that covers methods for suppressing intensity noise with photonic integrated noise eaters, which relates to the “Intensity Noise” section of this manuscript. The other authors do not have a competing interest.

%%%%%%%%%%%%%%%%%%%%%%% References %%%%%%%%%%%%%%%%%%%%%%%%%

%%%%%%%%%% If using BibTeX:
\bibliography{sample}

%%%%%%%%%%%%%%%%%%%%%%%%%%%%%%%%%%%%%%%%%%%%%%%%%%%%%%%%%%%%%%%
% Appendices
\section{Supplementary Information}
%%%%%%%%%%%%%%%%%%%%%%%%%%%%%%%%%%%%%%%%%%%%%%%%%%%%%%%%%%%%%%%

In the supplementary information we derive noise contributions related to squeezer saturation, phase noise, and intensity noise that are presented in the main text. We also include some background equations related to relative intensity noise and electronic noise.

%%%%%%%%%%%%%%%%%%%%%%%%%%%%%%%%%%%%%
\subsection{Saturation Bound}
\subsubsection{Nonresonant} \label{app:pump_depl_nonres}

We start with the analytic solution for the evolution of squeezed quadrature variance in lossless type-I phase-matched optical parametric amplification from \cite{LiKumar1995}
\begin{align}
    \hat y_1 (\zeta) &= f_1^-(\zeta) \hat y_1 (0) - f_3^-(\zeta)\hat y_3(0), \\
    V_1^- &= |f_1^-(\zeta)|^2 V_1^-(0) + |f_3^-(\zeta )|^2 V_3^-(0),
\end{align}
where $\hat y_i$ are the phase quadrature fluctuation operators for the squeezed field ($i=1$) and the pump field ($i = 3)$ and we assume the two operators are independent to combine simply and in terms of the corresponding squeezed quadrature variances $V_i^-$. The mixing coefficients evolve with normalized nonlinear length $\zeta = gL$ according to 
\begin{align}
   f_1^-(\zeta) & =  \text{sech}\zeta_0 \bigg[ \tanh\zeta_0\frac{\tanh(\zeta + \zeta_0)}{\text{sech}(\zeta + \zeta_0)} + \text{sech}(\zeta+\zeta_0)\big(1+\zeta \tanh\zeta_0 \big)\bigg], \\ 
   f_3^-(\zeta) &= -\frac{1}{\sqrt2}\text{sech}\zeta_0 \bigg[ \text{sech}\zeta_0 \frac{\tanh(\zeta + \zeta_0)}{\text{sech}(\zeta + \zeta_0)} + \text{sech}(\zeta+\zeta_0)\big(\zeta \text{sech}\zeta_0 -\frac{\tanh(\zeta_0)}{\text{sech}(\zeta_0)}\big)\bigg],
\end{align}
or in terms of mean field evolution
\begin{align}
   f_1^-(\zeta) & =  u_1(0) \bigg[ u_3(0)\frac{u_3}{u_1} + u_1\big(1-\zeta u_3(0)\big)\bigg], \\ 
   f_3^-(\zeta) &= -\frac{1}{\sqrt2} u_1(0)\bigg[ u_1(0)\frac{-u_3}{u_1} + u_1\big(\zeta u_1(0) +\frac{u_3(0)}{u_1(0)}\big)\bigg],
\end{align}
where the mean field amplitudes are normalized to the total input power $P_0 = P_1 + P_3$
according to $u_i(\zeta) = \sqrt \frac{P_i(\zeta)}{P_0}$
and evolve by \begin{align}
    u_1(\zeta) &= \text{sech}(\zeta + \zeta_0)\\
    u_3(\zeta) &= -\tanh(\zeta+\zeta_0)
\end{align}
and the initial conditions for power in the pump and seed are given by $\zeta_0$.
Under the typical OPA conditions of a strong undepleted pump $ u_3 \approx u_3(0) \approx 1, u_1(0)\ll 1$ and high gain $u_1(\zeta) \gg u_1(0)$, and defining $P_1^+$ as the portion of the seed along the amplified (phase) quadrature, we find 
\begin{align}
    |f_1^y |^2 &\approx \bigg(\frac{u_1(0)}{u_1}\bigg)^2 = \frac{P_1^+(0)}{P_1^+} \approx e^{-2gL} = \frac{1}{G} \\
    |f_3^y|^2 &\approx  \bigg(-\frac{1}{\sqrt2} {u_1}\bigg)^2 = \frac{P_1^+}{2P_0} \approx \frac{1}{2}G \frac{P_1^+(0)}{P_3(0)}.
\end{align}
and so we arrive at equation \eqref{eq:sat_deriv_start} in the main text 
\begin{align}
    V_1^-\approx \frac{P_1^+(0)}{P_1^+}V_1^-(0) + \frac{P^+_1}{2 P_0} V_3^-(0) \\
= V_\text{gain} + \Delta V_\text{pump saturation}
\end{align}

The minimum bound and optimal gain at this bound are derived from this:
\begin{align}
    V \approx \frac{1}{G} + \frac{G P_1^+(0)}{2P_0}V_3^Y\\
    G_{opt} = \sqrt{\frac{2P_0}{P_1^+(0)V_3^Y}}\\
    \Delta V_{min} = \sqrt{\frac{2P_1^+(0)V_3^Y}{P_0}}
\end{align}

%%%%%%%%%%%%%%%%%%%%%%%%%%%%%%%%%%%%%%%%
\subsubsection{Signal-resonant}
    \label{app:signal-res_sat}

The quadrature variances output from a signal-resonant squeezer are reported in \cite{Lam2004} and consist of the summation of numerous contributions to noise. In this section we show that the contribution from pump fluctuations simplifies to the ratio of signal to pump powers reported in the main text. 

The contribution from pump fluctuations to the anti-squeezed and squeezed quadratures are given by
\begin{align} \label{eq:opo_quads_refeq}
    V^\pm_\text{pump depl}
    &= \frac{\alpha^2 4 \kappa_{\mathrm{a}}^{(e)} \kappa_b^{(e)}(\epsilon/\kappa_b)^2 V_3^\pm}{|\kappa_a+\frac{[3;1]\epsilon^2 \alpha^2}{2 \kappa_b} \mp \epsilon \beta |^2},
\end{align}
with intracavity signal and pump fields (square root of photon number) $\alpha$ and $\beta$, total and extrinsic pump loss rates $\kappa_b$ and $\kappa_b^{(e)}$, total and extrinsic signal loss rates $\kappa_a$ and $\kappa_a^{(e)}$, and nonlinear coupling $\epsilon$.
The threshold pump field $\beta_{th}$ is given by
\begin{align}
    \kappa_a+\frac{3\epsilon^2 \alpha^2}{2 \kappa_b} - \epsilon \beta_{th}  = 0
\end{align}
and below threshold we have the bound
\begin{align}
    0<\frac{\epsilon^2 \alpha^2}{ \kappa_b} < \frac{2}{3}\epsilon \beta_{th} 
\end{align}
which implies
\begin{align} \label{eq:denom_bound}
    \epsilon(\beta + \frac{1}{3}\beta_{th}) < \kappa_a+\frac{\epsilon^2 \alpha^2}{2 \kappa_b} + \epsilon \beta_{}  < \epsilon(\beta + \beta_{th}).
\end{align}
We convert from photon number to power using
\begin{align} \label{eq:photonnum_to_power}
    \alpha^2 \kappa_a^{(e)} = \frac{P}{\hbar \omega}, \\
    \beta^2 \kappa_b = \frac{4 P_3}{\hbar 2\omega},
\end{align}
where $\hbar \omega$ is the energy of a signal photon. The factor of 4 in the pump equation arises from treating a traveling wave pump ($\kappa_b = \kappa_b^{(e)}$) in this resonant model (see SI of \cite{Stokowski2024}) and the factor of two in the denominator arises from the fact that a pump photon has twice the energy of a signal photon. 

Combining equations \ref{eq:opo_quads_refeq} - \ref{eq:photonnum_to_power} and using $\sqrt\frac{P_3}{P_{3,th}} = \frac{g}{g_{th}}$, we arrive at
\begin{align}
    V^-_{depl} > \frac{4}{(1 + {\frac{g}{g_{th}}})^2} \frac{ P_1 }{ 2 P_{3,th}} V_{3}^- ,  \\
    V^-_{depl} < \frac{4}{(1 + \frac{1}{3}{\frac{g}{g_{th}}})^2} \frac{ P_1 }{ 2 P_{3,th}} V_{3}^- ,  \\
\end{align}
and because $\frac{g}{g_{th}}<1$ below threshold, we arrive at the equations in the main text.

%%%%%%%%%%%
\subsubsection{Optical Parametric Generation}
The power in the squeezed vacuum is 
\begin{align}
        P_\text{SV} = \frac{G-\frac{1}{G}}{4}h\nu\Delta\nu.
\end{align}

Asserting that this power is much less than the pump power, we arrive at the loose bound given in the main text 
\begin{align}
    P_3 \gg \frac{G}{4}h\nu\Delta\nu, \\
    \frac{1}{G} \ll \frac{h\nu\Delta\nu}{4P_3},\\
    V \ll \frac{h\nu\Delta\nu}{4P_3}.
\end{align}

%%%%%%%%%%%%%%%%%%%%%%%%%%%%%%%%%%%
\subsection{Phase Noise Bound}
Here we derive the phase noise bound and optimal gain discussed in Section "Phase Noise". % \ref{sec:phasenoise}. 
Root-mean-square phase noise $\theta$ mixes the squeezed and antisqueezed quadratures according to \cite{Furusawa2006} 
\begin{align}
    V = \cos^2(\theta)V^- + \sin^2(\theta) V^+.
\end{align}
We use the Heisenberg uncertainty relation $V^- V^+ \geq 1 $ to bound the resulting squeezing in the presence of phase noise
\begin{align}
    V \geq \cos^2(\theta)V^- + \sin^2(\theta) \frac{1}{V^-}.
\end{align}
Optimizing over squeezer gain values and assuming a minimum uncertainty state, we find the optimal squeezing and gain are given by
\begin{align}
    V^-_\text{opt} &= \tan\theta , \\
    G_\text{opt} &= \cot\theta
\end{align}
and the resulting bound on measured variance is 
\begin{align}
    V\geq \sin 2\theta.
\end{align}

%%%%%%%%%%%%%%%%%%%%%%%%%
\subsubsection{Laser Phase Noise and Path Mismatch}

The general equation for phase error given the phase noise power spectral density $S_{\phi\phi}$ and path-length mismatch $\Delta L$ in material with group velocity $v_g$ is given by  
\begin{equation}
    \delta\theta_\text{laser} = \sqrt{ 4\times \frac{1}{2\pi}\int^\infty_{-\infty}S_{\phi\phi}(\omega)\sin^2(\frac{\omega \frac{|\Delta L|}{v_g}}{2})d\omega}.
\end{equation}
Assuming a laser with full-width at half-maximum linewidth $\Delta\nu$ and white frequency-noise spectrum \cite{Domenico2010}, the phase error simplifies to
\begin{equation}
\delta\theta_\text{laser} = \sqrt{2\pi \Delta\nu \frac{|\Delta L|}{v_g}}.
\end{equation}

%%%%%%%%%%%%%%%%%%%%%%%%
\subsubsection{Phase Locking Error} \label{app:phaselock}

Consider a simple phase locking setup consisting of a phase shifter that adjusts the relative phase between the signal and reference beams based on the error signal generated by control electronics, proportional to the difference in phase from the target phase. Let the system have input phase $\phi_\text{in}$, correction phase $\phi_\text{corr}$, and detected phase $\phi_\text{det} = \phi_\text{in} - \phi_\text{corr}$. In the below derivation we assume the locking bandwidth is much larger than the frequency-scales of the input phase drift, so that the lock is able to compensate for the drift.

The error signal $s$ is proportional to the detected phase drift 
\begin{align}
    s \approx K_\text{det} \times \phi_\text{det} + n 
\end{align}
where $K_\text{det}$ is the constant of proportionality and $n$ is the added measurement noise, with variance proportional to measurement bandwidth. 
The open-loop (no feedback) signal-to-noise-ratio of the error signal is \begin{align}
    \text{SNR}_0 = \frac{K_\text{det}^2 \delta\phi_\text{in}^2}{\delta n^2},
\end{align}
where $\delta \phi_\text{in}$ is the standard deviation of input phase error and $\delta n$ is the standard deviation of noise.
The correction phase is linearly proportional to the error signal
\begin{align}
    \phi_\text{corr} = K_\text{fb} s = K_\text{fb} K_\text{det} \times \phi_\text{det} + K_{fb} n,
\end{align}
which contains both the true feedback but also added noise. 
The resulting phase error can be solved from 
\begin{align}
    \phi_\text{det} &= \phi_\text{in} - \phi_{corr} \\
    & = \phi_\text{in} - K_\text{fb} K_\text{det} \times \phi_\text{det} - K_{fb} n, \\
    \phi_\text{det} &=  \frac{\phi_{in} - K_{fb}n}{1+K_{fb}K_\text{det}}     \\
    \delta\phi_\text{det}^2 &= \frac{\delta\phi_{in}^2 + K_{fb}^2\delta n^2}{(1+K_{fb}K_\text{det})^2} ,
\end{align}
where the noises are assumed to be independent and so their variances add separately.
Defining the loop gain $L_0 = K_{fb}K_\text{det}$ and using the open-loop SNR derived above, we arrive at the equation in the main text
\begin{align}
    \delta\phi_\text{det}^2 = \delta\phi_\text{in}^2 \frac{1+\frac{L_0^2}{\text{SNR}_0}}{(1+L_0)^2}, 
\end{align}
so in the low and high loop-gain limits 
\begin{align}
    \delta\phi^2 \approx \begin{cases}
	\delta\phi_\text{in}^2, \quad L_0 	\ll1 \\
	\frac{\delta \phi_\text{in} ^2}{\text{SNR}_0}, \quad L_0 \gg 1.
\end{cases}
\end{align}
Note in the high loop-gain limit the phase error is directly proportional to the noise from phase detection $\delta \phi _{L_0 \gg1} \propto \delta n$.

%%%%%%%%%%%%%%%%%
\subsection{Intensity Noise - Measurement Saturation}

We derive the contributions to noise that arise when the squeezed field is not vacuum but a displaced squeezed state. We find that for both balanced homodyne detection and parametric homodyne detection, the noise of the reference beam imprints on the squeezed signal if the power in the squeezed signal is non-negligible compared to the power in the reference beam.

\subsubsection{Balanced Homodyne Detection}

See, for example, Reference \cite{Bachor2019} for more background on modeling balanced homodyne detection. 
Consider the case of perfectly balanced homodyne with a small amount of leakage field $\hat a$ in the squeezed vacuum path and strong local oscillator $\hat b$ in the other port of the beamsplitter. 

\begin{align}
\hat a &= A + \hat{\delta X}_a + i\hat{\delta Y}_a,\\
\hat b &= \left[B + \hat{\delta X}_b + i\hat{\delta Y}_b\right]e^{i\phi_b}.
\end{align}

The detected fields are
\begin{align}
\hat c &= \frac{1}{\sqrt 2}(\hat a + \hat b),\\
\hat d &= \frac{1}{\sqrt 2}(\hat a - \hat b).
\end{align}

and the detected photocurrents are
\begin{align}
\hat N_c &= \frac{1}{2}\left(|\hat a|^2 + \hat b^\dag \hat a + \hat a^\dag \hat b + |\hat b|^2 \right),
\end{align}

with
\begin{align}
|\hat a|^2 &= \hat a^\dag \hat a \approx A^2 + 
A^\star \frac{\hat{\delta X}_a + i\hat{\delta Y}_a}{2} + 
A \frac{\hat{\delta X}_a - i\hat{\delta Y}_a}{2}
= A^2 + A \hat{\delta X}_a,\\
|\hat b|^2 &= \hat b^\dag \hat b = B^2 + B\hat{\delta X}_b.
\end{align}

Next,
\begin{align}
\hat b^\dag \hat a &=
\frac{2B^*A + \hat{\delta X}_b A - i\hat{\delta Y}_b A}{2}
+ \frac{2B^*\hat{\delta X}_a + \hat{\delta X}_b\hat{\delta X}_a - i\hat{\delta Y}_b \hat{\delta X}_a }{4} \\
&\quad + \frac{2 B^* i\hat{\delta Y}_a + \hat{\delta X}_b i\hat{\delta Y}_a - i\hat{\delta Y}_b i\hat{\delta Y}_a}{4} \\
&\approx B^*A + \frac{ \hat{\delta X}_b A - i\hat{\delta Y}_b A + B^*\hat{\delta X}_a + B^*i\hat{\delta Y}_a}{2},
\end{align}

and similarly,
\begin{align}
\hat a^\dag \hat b &\approx A^*B + \frac{ \hat{\delta X}_a B - i\hat{\delta Y}_a B + A^*\hat{\delta X}_b + A^* i\hat{\delta Y}_b }{2}.
\end{align}

We define the relative-phase quadratures:
\begin{align}
X_b &= B^* = B = |B|, \\
X_a &= A + A^*, \\
Y_a &= \frac{A - A^*}{i}.
\end{align}

Then the photocurrent becomes
\begin{align}
\hat N_c &\approx \frac{1}{2}\Big[
A^2 + A\hat{\delta X}_a + B^2 + B\hat{\delta X}_b 
+ \left(B X_a + \hat{\delta X}_a B + X_a\hat{\delta X}_b + Y_a \hat{\delta Y}_b\right)
\Big],\\
\hat N_d &= \frac{1}{2}\Big[
A^2 + A\hat{\delta X}_a + B^2 + B\hat{\delta X}_b 
- \left(B X_a + \hat{\delta X}_a B + X_a\hat{\delta X}_b + Y_a \hat{\delta Y}_b\right)
\Big].
\end{align}
The quadratures are defined with respect to the real local oscillator; $X_a^2 = P_1^-$, $Y_a^2 = P_1^+$, and $B^2 = P_{LO}$.
Differencing the two photocurrents cancels their average values and most intensity noise, leaving
\begin{align}
\hat N_c - \hat N_d &= 
B X_a + B\hat{\delta X}_a + X_a\hat{\delta X}_b + Y_a\hat{\delta Y}_b.
\end{align}
Thus the variance is
\begin{align}
V(\hat N_c - \hat N_d) = 
B^2 V(X_a)
+ B^2 V(\hat{\delta X}_a)
+ X_a^2 V(\hat{\delta X}_b)
+ Y_a^2 V(\hat{\delta Y}_b),
\end{align}
which in shot noise units becomes
\begin{align}
V_{\mathrm{BHD}} &= V(\hat{\delta X}_a) + V(X_a) + \frac{X_a^2 V(\hat{\delta X}_b) + Y_a^2 V(\hat{\delta Y}_b)}{B^2}, \\
V_{\mathrm{BHD}} &= V^- + \frac{P_1^- V^X_{\mathrm{LO}} + P_1   ^+ V^Y_{\mathrm{LO}}}{P_{LO}}.
\end{align}
If the LO is shot-noise-limited, the leakage penalty is simply the ratio $P_1/P_{LO}$. If the LO has excess noise, the penalty increases and depends on the amplitude/phase noise distribution and on the probe phase.

%%%%%%%%%%%
\subsubsection{Parametric Homodyne Detection} \label{app:PHD intensity noise}
In this section, we calculate the transfer of intensity noise from the pump to the signal in parametric homodyne detection.
The total seed power before and after the amplifier is
\begin{align}
    P_1(0) = \langle X \rangle ^2 + \langle Y \rangle ^2 + \frac{h\nu\Delta\nu}{4} (V^X + V^Y - 2), \\
    P_1 = G\langle X \rangle ^2 + \frac{1}{G}\langle Y \rangle ^2 + \frac{h\nu\Delta\nu}{4} (GV^X + \frac{1}{G}V^Y - 2),
\end{align}
for amplifier gain $G$, photon energy $h\nu$, optical squeezing bandwidth $\Delta \nu$, amplitude and phase quadratures $X, Y$ with variances $V^X, V^Y$ \cite{Bowen_qtm_metrology_bio}.
For simplicity, we define 
\begin{align}
    P_1^+ &= \langle X \rangle ^2 + \frac{h\nu\Delta\nu}{4}V^X, \\
    P_1^- &= \langle Y \rangle ^2 + \frac{h\nu\Delta\nu}{4}V^Y. 
\end{align}
Since parametric homodyne detection operates in the regime of macroscopic output powers, we neglect the microscopic term $\frac{h\nu\Delta\nu}{4}\times(-2)$ (though this could easily be added back in) and approximate the total signal power as $P_1 \approx P_1^+ + P_1^-$.
We linearize the power-dependent quadrature gain \(G(P)\) around the operating pump power \(P_3\):
\[
G \approx G_0 + G'\,\delta P_3, \quad G' \equiv \left.\frac{\partial G}{\partial P}\right|_{P_{3}} .
\]
The measured signal power is
\begin{align}
P_{\mathrm{det}} &= GP_{1}^+(0) + \frac{1}{G}P_1^-(0)\\
&= \big(G_0 + G'\,\delta P_3\big) P_{1}^+(0) + \frac{1}{(G_0 + \frac{\partial G}{\partial P} \delta P )}P_1^-(0) \\
&\approx (G_0 + \frac{\partial G}{\partial P} \delta P )P_1^+(0) + (\frac{1}{G_0 }- \frac{\frac{\partial G}{\partial P}}{G_0^2} \delta P)P_1^-(0). 
\end{align}
The mean detected power and fluctuations are therefore
\begin{align}
\langle P_{\mathrm{det}}\rangle &= P_1 = G_0 P_{1}^+(0) + \frac{1}{G_0}P_1^-(0),\\
\delta P_{\mathrm{det}} &= G'\,\delta P_3 \bigg( P_{1}^+(0) - \frac{1}{G_0^2}P_1^-(0)\bigg) .
\end{align}
Hence the variance of the detector power arising from pump-power noise is
\begin{align}
\delta P_{\mathrm{det}}^2
&= \bigg(G'\big( P_{1}^+(0) - \frac{1}{G_0^2}P_1^-(0)\big)\bigg)^2\,\delta P_3^2.
\end{align}
Converting from power variance to shot noise units, 
\begin{align}
    V_{\mathrm{det}}  &= \frac{\delta P_\text{det}^2}{P_\text{det}2h\nu}, \quad 
    V_{\mathrm{3}}^X  = \frac{\delta P_\text{3}^2}{P_\text{3}2h(2\nu)}\\
    V_{\mathrm{det}}   &= 2(G')^2  \frac{\bigg(P_1^+(0) -\frac{1}{G_0^2}P_1^-(0)\bigg)^2}{ \langle P_\text{det}\rangle } P_3  V_3^X(0) .
\end{align}
Note that the goal of PHD is to amplify the measurement quadrature sufficiently such that most of the detected power comes from this amplified quadrature $G_\text{det}P_1^+(0) \gg \frac{P_1^-(0)}{G_\text{det}}$.
Therefore we can simplify the above expression to
\begin{align}
    \Delta V_{\mathrm{det}}   &= 2(G')^2  \frac{\bigg(P_1^+(0) \bigg)^2}{G_\text{det} P_1^+(0) } P_3  V_3^X(0) \\
    &=  2(G')^2  \frac{P_1^+(0)}{G_\text{det} } P_3  V_3^X(0),  
\label{eq:Vdet_V3}
\end{align}
which is the equation in the main text.

Now we can use this result to calculate the transduced intensity noise from the pump to the signal in specific squeezer designs.
For a nonresonant squeezer, we have 
\begin{align}
   G' =  \frac{\partial G}{\partial P}|_{P_3} = \frac{\partial (e^{2\sqrt{\eta P} L})}{\partial P}|_{P_3} = \frac{2\sqrt\eta L}{2 \sqrt P_3} e^{2gL} = \frac{\sqrt\eta L}{ \sqrt P_3} G_0 = \frac{gL}{P_3}G_0\\
\end{align}
so we get
\begin{align}
    \Delta V_\text{det}  = 2(gL)^2\frac{G_0 P_1^+(0)}{P_3 } V_3^X(0).
\end{align}

Similarly, for a signal resonant squeezer we have 
\begin{align}
    G &= V^+= 1+\rho \frac{4 {g}/{g_{\mathrm{th}}}}{(1-g/g_{\mathrm{th}})^2},\\
    \frac{\partial G}{\partial P}|_{P_3} &= \frac{2\rho\sqrt{\eta_0} L g_{\mathrm{th}} (g_{\mathrm{th}}+g)}{\sqrt{P_3} (g_{\mathrm{th}}-g)^3}
=
\frac{2\rho \frac{g}{g_{th}} (1+\frac{g}{g_{\mathrm{th}}})}{P_3 (1-\frac{g}{g_{\mathrm{th}}})^3}
\\
&\approx \frac{G_0}{P_3(1-\frac{g}{g_{\mathrm{th}}})} , \ \ \ \ \ (1-\frac{g}{g_{th}}) \ll 1, G \gg 1\\
\Delta V_{\mathrm{det}}   &=  2\frac{G_0}{(1-\frac{g}{g_{\mathrm{th}}}) )^2} \frac{ P_1^+(0)}{ P_3}   V_3^X(0) .
\end{align}

%%%%%%%%%%%%%%%%%%%%%%%%%%%%%%%%%%%%%%%
\subsection{Relative Intensity Noise in Shot Noise Units}
\label{app:RIN_sn}
The total (single-sided) power spectral density of light is the sum of the contributions from shot noise and classical noise (which add linearly, since they are uncorrelated)
\begin{align}
    S_{P} &= S_{P,sn} + S_{P,cl}
\end{align}
The power spectral density of shot-noise-limited light is
\begin{align} \label{eq:app_sn_PSD}
    S_{P,sn} &= 2 h\nu P,
\end{align}
where $h\nu$ is the photon energy and $P$ is the optical power.
The relative intensity noise of light is defined as the power fluctuations divided by the mean power squared
\begin{align}
    \mathrm{RIN} &= \frac{S_P}{P^2}\\
    &= \mathrm{RIN}_\text{sn} + \mathrm{RIN}_\text{cl}
\end{align}
and the intensity quadrature variance of light is 
\begin{align}
    V^X = \frac{S_P}{2h\nu P}
\end{align}
Rearranging the above equations, we arrive at the equations in the main text
\begin{align}
V^X &= \frac{S_{P,\text{sn}} + S_{P,\text{cl}}}{2h\nu P}\\
    &= \frac{2h\nu P}{2h\nu P} + \frac{\text{RIN}_\text{cl} P^2}{2 h \nu P} \\ 
    &=1 +  \frac{P}{P_{\text{RIN}_\text{cl=sn}}}, \ \ \ \ \text{where }P_{\text{RIN}_\text{cl=sn}} = \frac{2h\nu}{\text{RIN}_\text{cl}}\\
    & = V^X_\text{shot noise} + V^X_\text{classical} .
\end{align}

%%%%%%%%%%%%%%%%%%%%%%%%%%%%%%%%%%%%%%%
\subsection{Electronic Noise in Shot Noise Units}
\label{app:elec_sn}

The electronic noise of a photodetector contains contributions from the dark current's noise as well as the Johnson noise, and is often reported as a noise-equivalent power (NEP), which is the optical power that produces a signal-to-electronic-noise ratio of 1. 
Relative to shot noise (equation \eqref{eq:app_sn_PSD}), the electronic noise is
\begin{align}
    V_\text{elec} = \frac{\text{NEP}^2 \eta_\text{QE} }{2 h \nu} \frac{1}{P_\text{det}},
\end{align}
for detected power $P_\text{det}$ and detector quantum efficiency $\eta_\text{QE}$. % NOTE that NEP has a hidden internal factor of 1/eta_QE
The factor $\eta_\text{QE}$ appears in the numerator because the noise-equivalent power is referenced to optical power prior to photodetection, and therefore already scales inversely with detector quantum efficiency.

\end{document}